\PassOptionsToPackage{sort&compress}{natbib}
\documentclass{WileyMSP-template}
\usepackage{eqnarray,amsmath,physics,bbold}
\usepackage{subfig}
\usepackage{physics}
\usepackage{tabularx}
\usepackage{xspace}
\usepackage{textcomp}
\usepackage{mathtools}
\usepackage{xcolor}
\usepackage{hyperref}
\usepackage{svg}
\usepackage{natbib}
\setcitestyle{super,comma,square}
\usepackage{comment}

\newcommand{\ped}[1]{_{\text{#1}}}
\newcommand{\api}[1]{^{\text{#1}}}

\begin{document}

\pagestyle{fancy}
\rhead{\includegraphics[width=2.5cm]{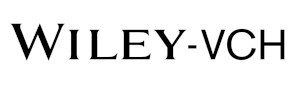}}

\title{Mitigating Errors on Superconducting Quantum Processors through Fuzzy Clustering}

\maketitle

\author{Halima G. Ahmad*, Roberto Schiattarella, Pasquale Mastrovito, Angela Chiatto, Anna Levochkina, Martina Esposito, Domenico Montemurro, Giovanni P. Pepe, Alessandro Bruno, Francesco Tafuri, Autilia Vitiello, Giovanni Acampora, Davide Massarotti}

\begin{affiliations}

H.G. Ahmad, P. Mastrovito, A. Levochkina, G.P. Pepe\\
Dipartimento di Fisica ``Ettore Pancini'', Università degli Studi di Napoli Federico II, Via Cinthia, I-80126, Napoli, Italy\\
CNR-SPIN, UOS Napoli, Monte S. Angelo, via Cinthia, I-80126, Napoli, Italy\\

R. Schiattarella, A. Chiatto, D. Montemurro, A. Vitiello, G. Acampora\\
Dipartimento di Fisica ``Ettore Pancini'', Università degli Studi di Napol Federico II, Via Cinthia, I-80126, Napoli, Italy\\

A. Bruno\\
QuantWare, Elektronicaweg 10, 2628 XG Delft, The Netherlands\\

M. Esposito\\
CNR-SPIN, UOS Napoli, Monte S. Angelo, via Cinthia, I-80126, Napoli, Italy\\

F. Tafuri\\
Dipartimento di Fisica ``Ettore Pancini'', Università degli Studi di Napoli Federico II, Via Cinthia, I-80126, Napoli, Italy\\
Consiglio Nazionale delle Ricerche-Istituto Nazionale di Ottica (CNR-INO), Largo Enrico Fermi 6, I-50125 Florence, Italy\\

D. Massarotti\\
Dipartimento di Ingegneria Elettrica e delle Tecnologie dell’Informazione, Università degli Studi di Napoli Federico II, I-80125 Napoli, Italy\\
CNR-SPIN, UOS Napoli, Monte S. Angelo, via Cinthia, I-80126, Napoli, Italy\\

Email Address: halimagiovanna.ahmad@unina.it

\end{affiliations}

\keywords{Superconducting Quantum Computing, Quantum Error Mitigation, Fuzzy Clustering}

\begin{abstract}

Quantum utility has been severely limited in superconducting quantum hardware until now by the modest number of qubits and the relatively high level of control and readout errors, due to the intentional coupling with the external environment required for manipulation and readout of the qubit states. Practical applications in the Noisy Intermediate Scale Quantum (NISQ) era rely on Quantum Error Mitigation (QEM) techniques, which are able to improve the accuracy of the expectation values of quantum observables by implementing classical post-processing analysis from an ensemble of repeated noisy quantum circuit runs. In this work, we focus on a recent QEM technique that uses Fuzzy C-Means (FCM) clustering to specifically identify measurement error patterns. For the first time, we report a proof-of-principle validation of the technique on a $2$-qubit register, obtained as a subset of a real NISQ $5$-qubit superconducting quantum processor based on transmon qubits. We demonstrate that the FCM-based QEM technique allows for reasonable improvement of the expectation values of single- and two-qubit gates-based quantum circuits, without necessarily invoking state-of-the-art coherence, gate, and readout fidelities.

\end{abstract}
\section{Introduction}
\label{sec:intro}

The growing development of quantum technologies has led the field closer to an era of quantum utility~\cite{arute2019,kim2023evidence}. By exploiting the peculiar principles of superposition and entanglement, quantum computing promises polynomial and super polynomial speed-up over the best-known classical algorithms for the resolution of different major problems~\cite{Shor_1997,feynman2018,pirandola2020advances, egger2020quantum, li2020quantum, acampora2022implementation, blunt2022perspective}. However, the actual applicability of these algorithms is severely limited by current quantum hardware capabilities. Indeed, up-to-date quantum systems are known as Noisy Intermediate Scale Quantum (NISQ) computers, due to the limited number of the available qubits and the relatively high level of errors, due to state-leackage, qubit-qubit cross-coupling and coupling with the external environment, which affects the computing power they can provide~\cite{preskill2018quantum}. While even a modest number of qubits is still sufficient for complex computations, on the other hand quantum and environmental noise severely limit the race toward quantum advantage. 

It is widely believed that to overcome this limitation, quantum computers must be equipped with Quantum Error Correction (QEC) techniques in the long-term~\cite{cai2023}. QEC is a very active research area in quantum engineering~\cite{Acampora2021,cai2023}, and proposes to encode quantum information in redundant ways to make it resilient to noise~\cite{nielsen1998,cai2023}. One possible solution for fault-tolerant quantum computing is, for example, to encode quantum information into aggregates of more than one physical qubit, known as logical qubits~\cite{nielsen1998,cai2023}. In this approach, data qubits must be coupled to auxiliary ancilla qubits in order to detect and correct quantum errors (surface codes)~\cite{Fowler2012,cai2023}. Surface-code fault-tolerant techniques require a multiplicative increase in the number of working qubits, with a significant overhead (from a few thousands to hundreds of thousands) in order to solve hard computational problems~\cite{Fowler2012,Kivlichan2020, Lee2021,cai2023}. Since this is currently not feasible in NISQ devices, alternative techniques known as \textit{Quantum Error Mitigation} (QEM) have been proposed, which can provide an immediate improvement of current experimentally implementable quantum algorithms~\cite{cai2023}.

Compared to QEC, QEM techniques have the ability to minimize noise-induced bias in expectation values or in distribution sampling on noisy hardware by implementing post-processing algorithmic schemes from an ensemble of circuit runs, thus not requiring further hardware resources~\cite{cai2023}. Indeed, they are able to identify, correct, and mitigate the types of errors that affect the output of a quantum algorithm, and therefore to improve the overall accuracy of the final outcomes~\cite{cai2023}. 

In the realm of QEM techniques, it's important to recognize the wide variety of approaches available. The Fuzzy C-Means-based QEM (FCM-QEM) technique proposed by Acampora and Vitiello in \cite{Acampora2021}, is a valuable method and primarily targets the correction of measurement errors. Differently from other techniques that require adjusting parameters or have a high sampling cost~\cite{cai2023}, like Zero-Noise Extrapolation (ZNE)~\cite{Li2017,temme2017}, Probabilistic readout and control Error Cancellation (PEC)~\cite{temme2017} or the
Twirled Readout Error Extinction (t-REx)~\cite{vandenberg2022}, the approach proposed in \cite{Acampora2021} requires only supplementary classical post-processing, being extremely convenient from an experimental point of view.
 
The main idea behind FCM-QEM is to use the soft clustering approach to identify different types of readout errors, and use this information to create a so-called mitigation matrix, which is able to transform quantum outcomes in order to reduce the errors. In~\cite{Acampora2021}, the FCM-QEM approach has been demonstrated to outperform error measurement mitigation methods available in the IBM Python package named Qiskit~\cite{Qiskit}. However, the experiments have only focused on mitigating readout errors on noisy quantum simulators with varying levels of quantum measurement noise. As a result, the FCM-QEM approach has yet to be applied to a NISQ device. To bridge this gap, we propose for the first time to apply the FCM-QEM technique on a real NISQ device based on superconducting transmon qubits~\cite{koch2007,krantz2019}. 

Superconducting Quantum Processing Units (s-QPUs) are so far the most promising quantum platforms for NISQ and fault-tolerant quantum computing~\cite{Fowler2012,gambetta2017,versluis2017,campbell2017,egan2021,krinner2022,webster2022,zhao2022,marques2022}. The quantum macroscopic circuital nature of these systems, which relies on the use of Josephson non-linear devices as fundamental building blocks~\cite{Barone1982,tafuri2019}, as well as superconducting transmission lines and resonant elements~\cite{krantz2019}, allows to engineer \emph{ad-hoc} on-chip solutions to comply on a certain extent with the DiVincenzo's criteria~\cite{DiVincenzo}, starting from the possibility to manipulate and control the qubit state~\cite{krantz2019,aguado2020,Ahmad2022,Bao2022}, and implement efficient qubit state readout protocols~\cite{koch2007,lecocq2021,DiPalma2023}. Indeed, Josephson-based s-QPUs have already shown remarkable capabilities that point towards quantum advantage~\cite{arute2019,kim2023}, by exploiting state-of-the-art QEM techniques~\cite{kandala2019,song2019,smith2021,vandenberg2022,kim2023}. We show that FCM-QEM promises similar performances. Specifically, we report a proof-of-concept validation of the technique on a $2$-qubit register, obtained as a subset of a $5$-qubit s-QPU. 

The paper is organized as follows. In Section~\ref{sec:error_mitigation}, we first dive into the main concepts behind error mitigation in quantum measurements and the FCM-QEM technique. In Section~\ref{sec:hardware}, we describe the hardware setup, starting from the s-QPU characterization, the coherence and average gate and readout fidelity benchmarking of the $2$-qubit register under analysis. The techniques employed for the experimental validation of the FCM-QEM approach are reported in Section~\ref{sec:exp_and_res}. In Section~\ref{sec:conclusion}, we discuss the final results and demonstrate that the FCM-QEM technique allows for a reasonable improvement of the readout, without necessarily invoking state-of-the-art coherence, gate, and readout fidelities for the hardware. Finally, we propose future experiments to possibly outline the performance of FCM-QEM on a scalable s-QPU system.

\section{Measurement Error Mitigation through Fuzzy Clustering}
\label{sec:error_mitigation}

\begin{figure}
    \centering
    \includegraphics[scale=0.6]{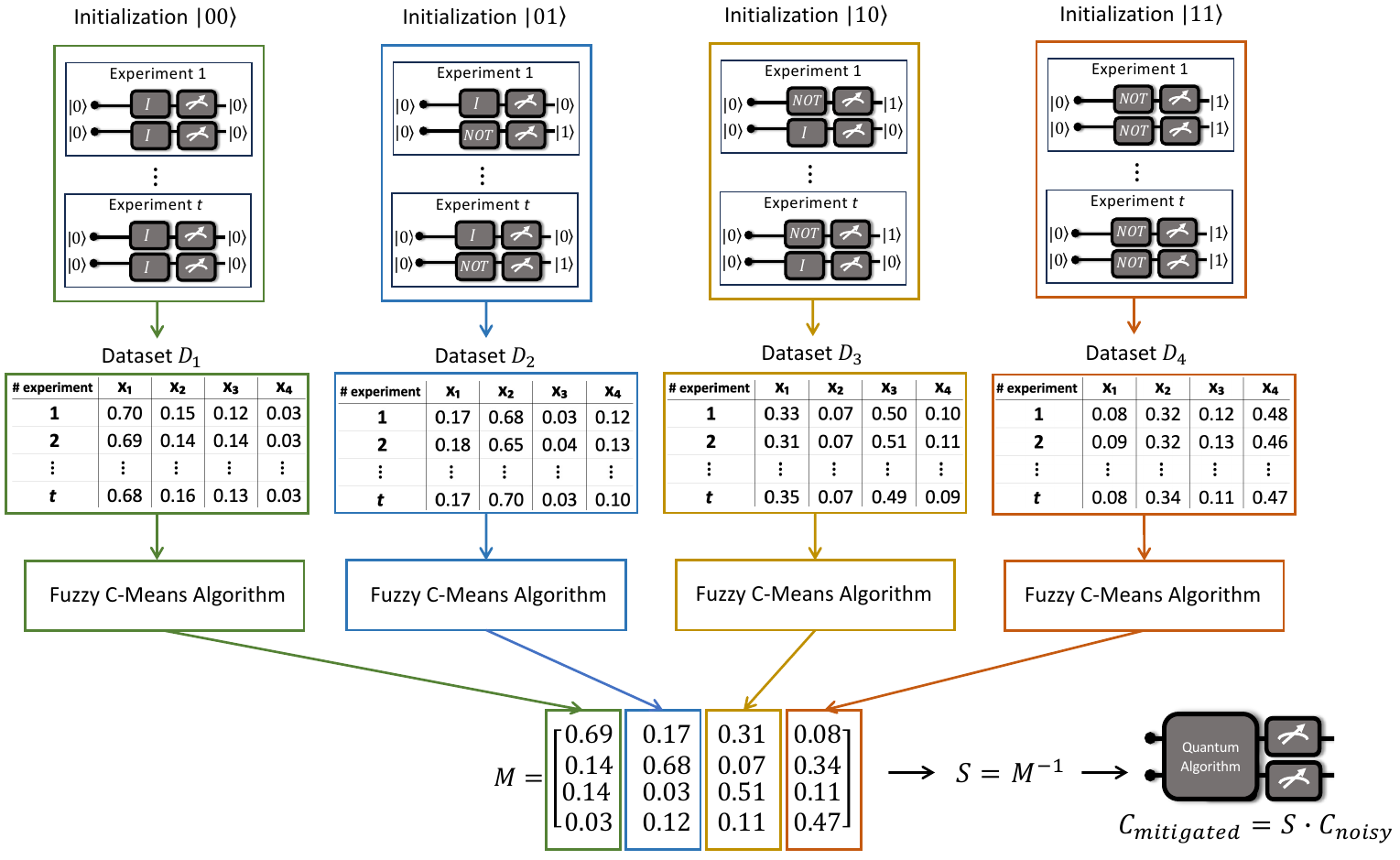}
    \caption{Graphical representation of the workflow of the FCM-QEM method for a $2$-qubit register. The first step consists in creating four different datasets, each of which is obtained by initializing the $2$-qubit register in one of all the 4 computational basis states $\{ \ket{00}, \ket{01}, \ket{10}, \ket{11}\}$ and running $t$ different experiments. Each dataset $D_i$ is composed of $t$ instances each described by ($x_1$, $x_2$, $x_3$, $x_4)$, which are the probabilities of measuring the basis states $\{ \ket{00}, \ket{01}, \ket{10}, \ket{11}\}$, respectively. The clustering of the data for each dataset uses the Fuzzy C-Means algorithm, and the calibration matrix $M$ is computed using the probability vectors belonging to each cluster with a similar membership degree in the fuzzy step. The mitigation matrix $S$ is obtained by inverting $M$, and can be used to mitigate the noisy count vectors $C_{noisy}$ of any quantum algorithm implemented on the quantum processor.}
    \label{fig: FCM-QEM}
\end{figure}

Noise in existing quantum hardwares inhibits the reliability of quantum computation, because it unavoidably introduces classical and quantum errors~\cite{preskill2018quantum}. The \textit{quantum measurement errors} are among the most relevant. State-of-the-art quantum hardware may experience readout errors ranging from $8\%$ to $30\%$~\cite{tannu2019mitigating}, which can occur as bit-flips, i.\,e., where outcomes are mistakenly recorded as $\ket{0}$ when they are $\ket{1}$, and \emph{vice versa}. Even if the quantum processor ideally performs noiseless operations, the final measurement introduces random perturbations, which makes the final $n$-bit string different from the ideal one. To address this computational error, QEM techniques can be employed as a post-processing step on a classical computer, aiming at rectifying the noisy outcome generated by the quantum device. One approach currently used in NISQ devices involves computing the \textit{mitigation matrix}, denoted as $S$, which is exploited to classically correct the noisy quantum result. In detail, let $C^r\ped{noisy}$ be a noisy count vector output of an $n$-qubit circuit run $r$ times. We can model the measurement error as a linear transformation of the ideal noise-free count vector  $C^r\ped{ideal}$ as~\cite{cai2023}:
\begin{equation}
    C^r\ped{noisy} = M \cdot C^r\ped{ideal},
\end{equation}
where $M$ is the \textit{calibration matrix}, defined by a $2^n \times 2^n$ matrix that describes the measurement errors of the quantum processor. Each column of the matrix $M$ represent a probability vector (\textit{i.e.} the count vector divided by the total number $r$ of shots) computed when the $n$-qubit circuit is prepared in one of the possible $2^n$ computational basis states. For instance, in the case of a $2$-qubit register, the calibration matrix $M$ is a $4 \times 4$ matrix, where the columns are the probability vectors obtained by initializing the register in the four computational basis states $\{\ket{00}$, $\ket{01}$, $\ket{10}$, $\ket{11}\}$ for $r$ shots, respectively. Considering the calibration matrix $M$, the classical mitigation approach consists in constructing the mitigation matrix as the inverse matrix $S=M^{-1}$, and then applying it to the count vector obtained from a specific quantum experiment. Formally, to convert the noisy output into the ideal one, the classical approach uses:
\begin{equation}
C^r\ped{ideal} = S \cdot C^r\ped{noisy}.
\end{equation}

This approach shows a critical drawback: the calibration matrix $M$ is not unique, but can change by calculating the matrix several times. An inaccurate computation of $M$ can be detrimental to the error mitigation technique, leading to a worse output than the mitigation-less case. 
To avoid this issue and pick the most suitable calibration matrix $M$, here we use the fuzzy-based approach introduced in ~\cite{Acampora2021}.

This approach uses a fuzzy clustering algorithm, called Fuzzy C-Means (FCM), to support an efficient selection of the mitigation matrix. The workflow of this method (see Figure \ref{fig: FCM-QEM}) is made of three main steps: 1) the dataset creation step, aimed at collecting the information about the behavior of the quantum processor involved; 2) the fuzzy step, aimed at applying the FCM algorithm and clustering the collected probability vectors in different error patterns; 3) the mitigation matrix computation step, aimed at selecting the most appropriate probability vectors to calculate the mitigation matrix.
To better understand how the FCM-QEM technique works, we can consider its application on a generic $n$-qubit register. The first step consists in creating a dataset collection containing information about the preparation of all the $2^n$ computational basis states in which the register can be prepared. For each computational basis state, we collect $t$ different probability vectors corresponding to the output of $t$ different experiments. Therefore, the $i$-th dataset $D_i$ ($i = 1,2,...,2^n$) is made up of $t$ instances, each consisting of $2^n$ elements, defined as $\textbf{x}_j = ( x_{j,1}, \dots , x_{j,{2^n}})$ ($j=1, \dots, t$). 
Roughly, the information contained in each dataset $D_i$ represents the behaviour of the quantum processor when it is initialized in any of the computational states. This means that similar count vectors in each dataset identify a similar error pattern for the quantum processor. Therefore, performing clustering procedure on these datasets allows detecting the different error patterns that affect the quantum processor. 

In the second step, FCM is actually applied to build $C$ clusters on each dataset $D_i$. In particular, FCM assigns to each instance $j$ a membership degree $w_{k_Cj}$ belonging to one of the clusters $k_C$. Following the algorithm proposed in~\cite{bezdek1984fcm}, this operation is performed by minimizing the cost function:
\begin{equation}\label{eq: J^FCM}
    J(W, V) = \sum_{k_C=1}^{C} \sum_{j=1}^{t} (w_{k_Cj})^m ||\textbf{x}_{j} - \textbf{v}_{k_C}||^2,
\end{equation}
where $W = w_{i_Cj}\in[0,1]$ defines the so-called fuzzy partition matrix, $V = \{\textbf{v}_1, \dots , \textbf{v}_C\}$ is the set of cluster centroids, and $m\in (1, + \infty)$ is the fuzziness coefficient (or fuzzifier), and $||\textbf{x}_{j} - \textbf{v}_{k_C}||^2$ is the Euclidean distance between the $k_C$-th cluster center and the $j$-th instance. The optimization of Eq.~\ref{eq: J^FCM}  returns the final partition matrix $W$ and the set of centroids $V$ when the maximum number of iterations $maxiter$ is reached, or the difference between two consecutive fuzzy partition matrices is less than a threshold value $\phi$. The value $C$ is set \emph{a priori} to run the FCM algorithm for each dataset $D_i$. The optimal value of $C$ is found by running the FCM algorithm over different values of $C$ and picking the one that maximizes the coefficient $fpc$, called fuzzy partition coefficient\cite{Acampora2021}. This coefficient, introduced in~\cite{ross2009fuzzy}, assumes values in the range $fpc\in[0, 1]$ where the higher $fpc$ is, the better the quality of the chosen set of clusters. The final result for each dataset $D_i$ will thus be a fuzzy partition matrix $W_i$ obtained from the best $C$ value among those tested.

The last step is to compute the calibration matrix $M$, and, as a consequence, the mitigation matrix $S$, by taking into account the most appropriate probability vectors obtained by quantum experiments executed on the quantum processor. In detail, the optimal probability vector is characterized by the most uncertain membership degree in the fuzzy partition matrix $W_i$. This choice is related to the fact that this probability vector can represent the best trade-off among the pattern errors since it belongs to each cluster identified in the fuzzy step with similar membership degrees. Finally, the calibration matrix $M$ obtained with these probability vectors is inverted to obtain the mitigation matrix $S$, which can be used to mitigate the noisy count vector $C\ped{noisy}$ of different quantum algorithms run on the quantum processor.

\section{Superconducting Quantum Processor: Characterization and Experimental Techniques}\label{sec:hardware}

\begin{figure}[t]	
\subfloat[][\centering]{\includegraphics[width=0.6\columnwidth]{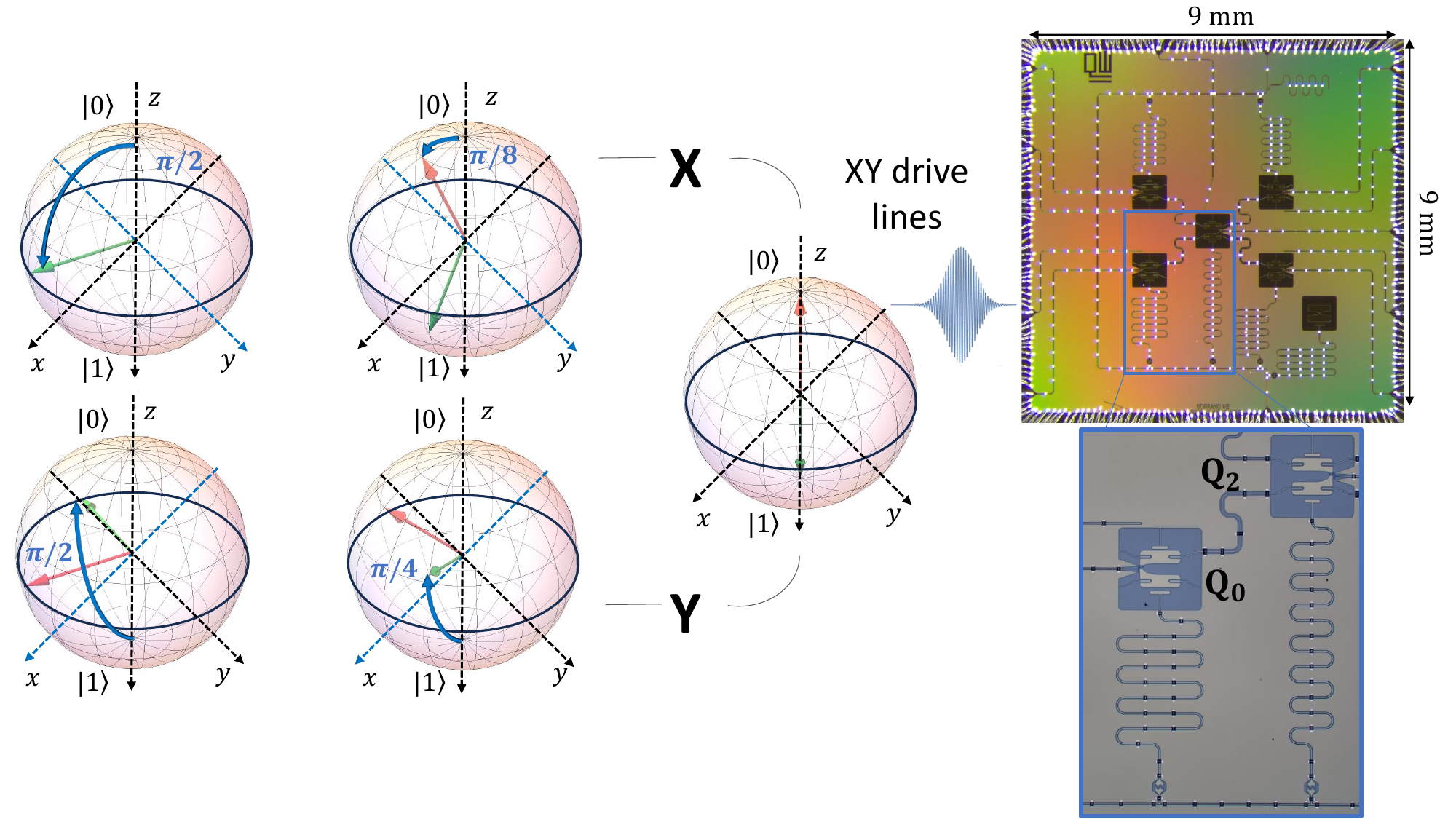}}
\subfloat[][\centering]{\includegraphics[width=0.45\columnwidth]{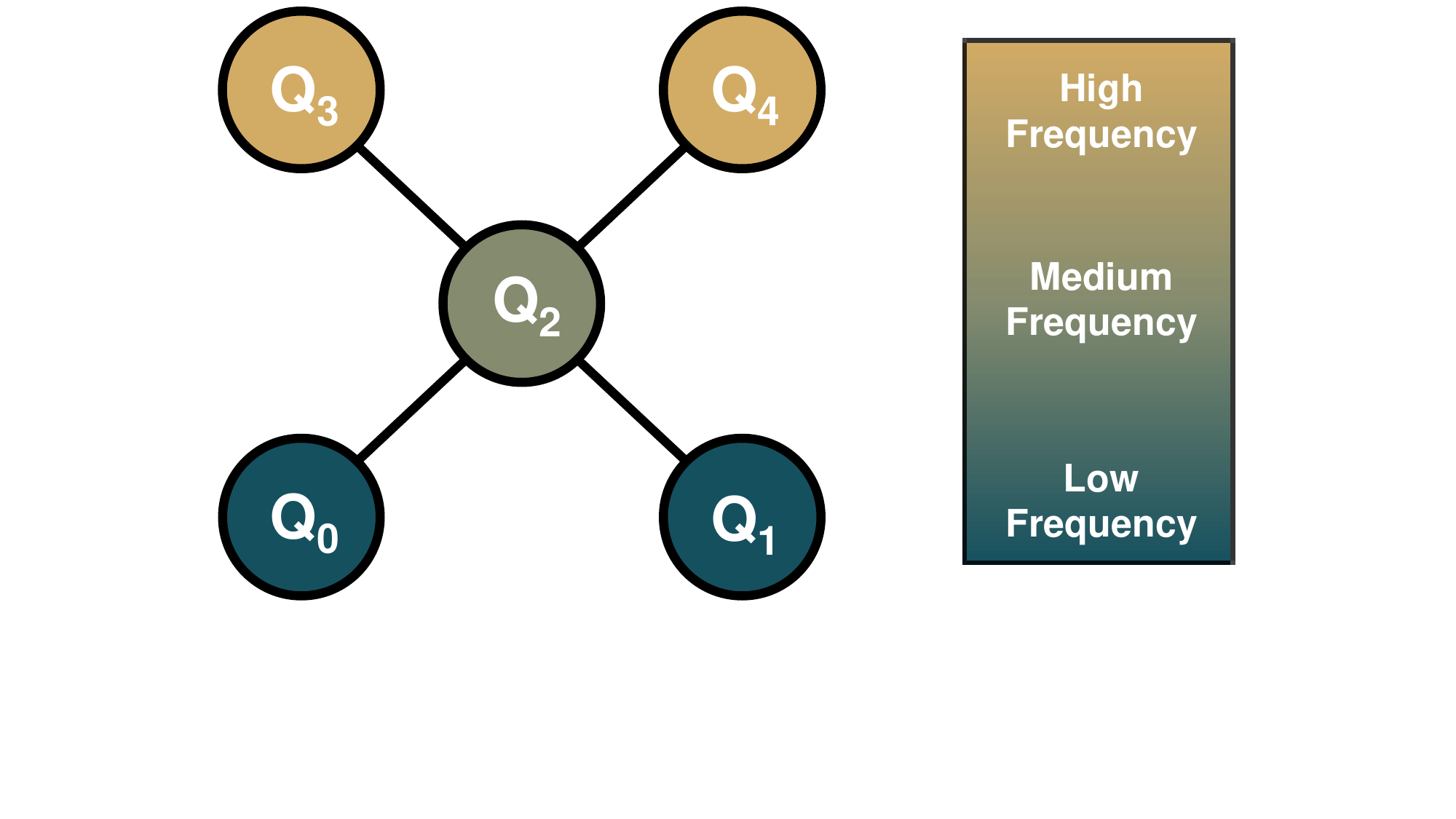}}
\centering
	\caption{In a), optical microscope of the $5$-qubit quantum processor used in this work. The implementation of the Fuzzy C-means quantum error mitigation technique specifically involved the pair $Q_0$-$Q_2$, highlighted in the blue box with larger magnification. Details on the materials and the circuital design can be found in ~\cite{Ahmad2023}. On the drive line of $Q_0$, we show a pictorical representation of the impact of microwave XY-pulses on the qubit state using the Bloch sphere representation. In b), schematic representation of the $5$-qubit chip frequencies: in dark green, two low-frequency flux-tunable transmon qubits (dark green) are coupled to a medium-frequency central flux-tunable transmon qubit (in light green) through high-frequency bus resonators. The central qubit is also coupled via resonator buses to two high-frequency flux-tunable transmon qubits (in yellow).} 
	\label{sample}     
\end{figure} 

To experimentally validate the performance of the FCM-QEM technique we employ a $5$ transmons s-QPU (\emph{Soprano}, manufactured by QuantWare), whose schematic is reported in Figure~\ref{sample}. The s-QPU is composed of two low-frequency qubits (in dark green), one central medium-frequency qubit (in light green) and two high-frequency qubits (in yellow). Full-connectivity between each qubit is achieved through four high-frequency bus resonators~\cite{Majer2007,Sillanpaa2007} between the central qubits and the other qubits of the quantum processor, while two-qubit connectivity is guaranteed between the low-frequency and high-frequency qubits through the central qubit. Details on the coupling bus resonators are reported in~\cite{Ahmad2023}. Each qubit is capacitively coupled to a dedicated microwave control line, i.e. to perform XY rotations~\cite{krantz2019}, and to a superconducting resonator for dispersive readout of the qubit state~\cite{koch2007,krantz2019}. Each readout resonator is capacitively coupled to a common feedline in a notch-type geometry, in order to guarantee multiplexing and simultaneous readout~\cite{George2017}. Moreover, the qubits in the circuits are flux-tunable transmons~\cite{koch2007}, i.e. it is possible to tune the qubit frequency by means of dedicated fast flux lines and to implement flux-based two-qubit gates~\cite{koch2007,krantz2019}. Details on the materials and the circuit design are reported in~\cite{Ahmad2023}. Additional details on the experimental setup employed in this work are collected in the Supplementary Material.

\begin{figure}
\centering
	\subfloat[][\centering]{\includegraphics[width=0.3\columnwidth]{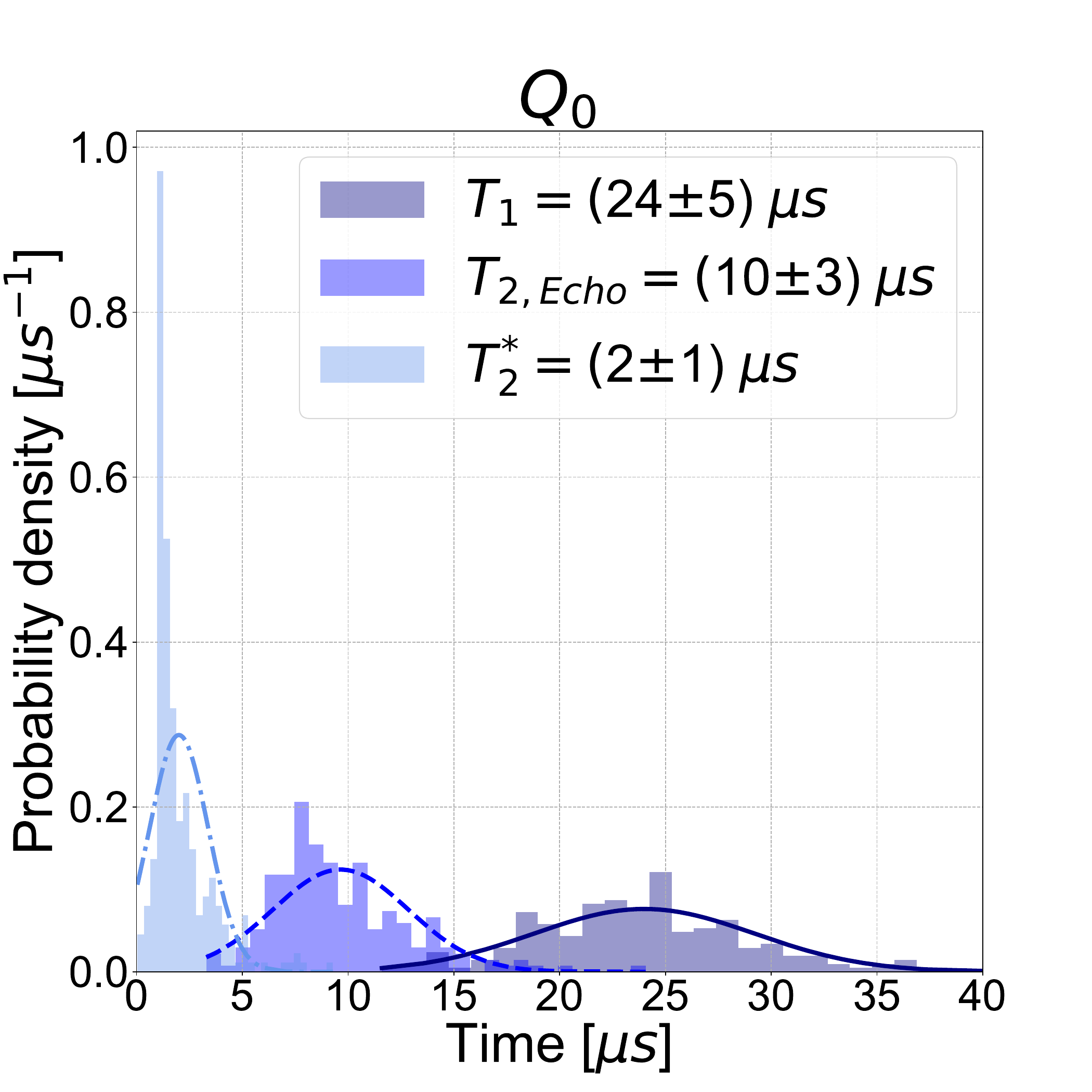}}
	\subfloat[][\centering]{\includegraphics[width=0.3\columnwidth]{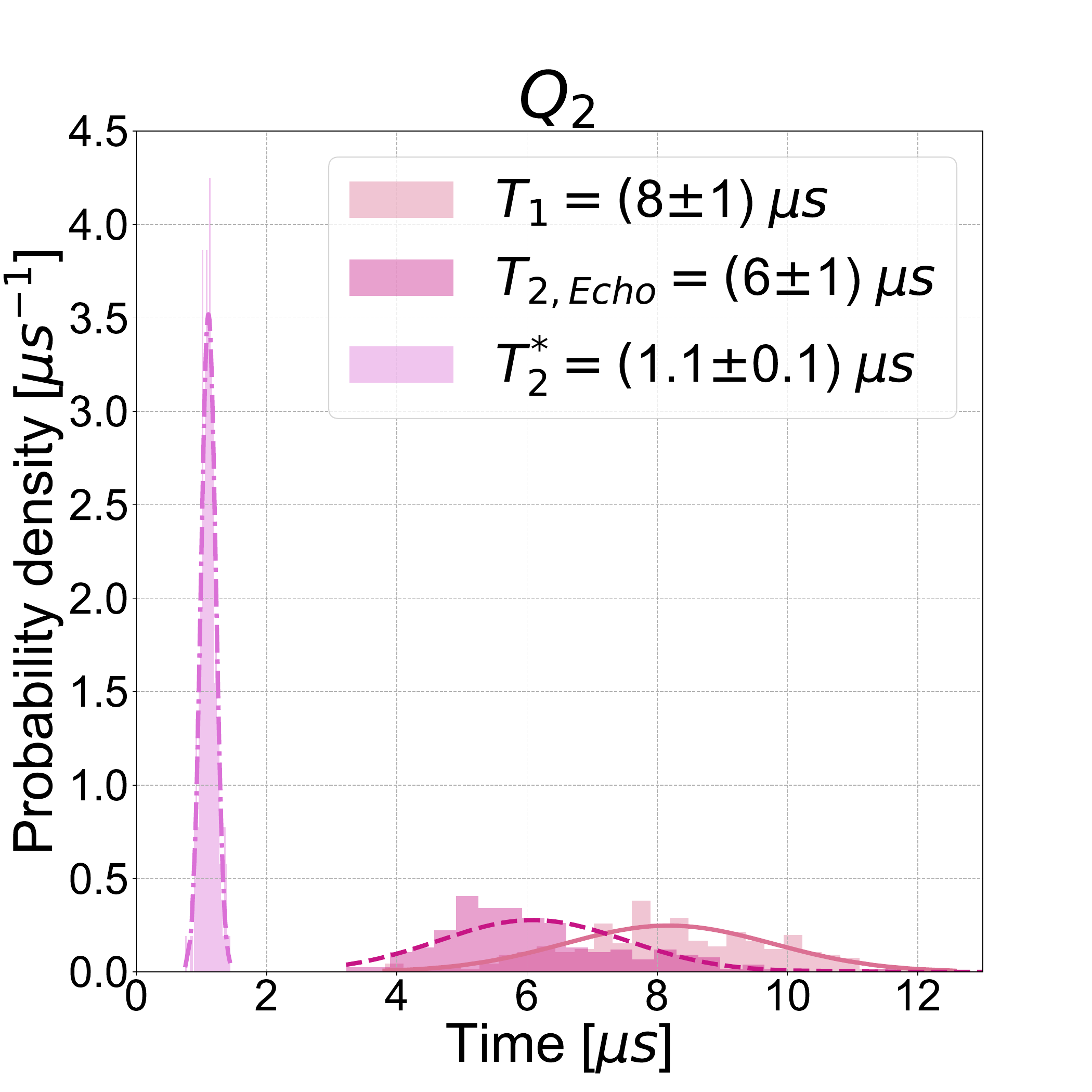}}
	\subfloat[][\centering]{\includegraphics[width=0.3\columnwidth]{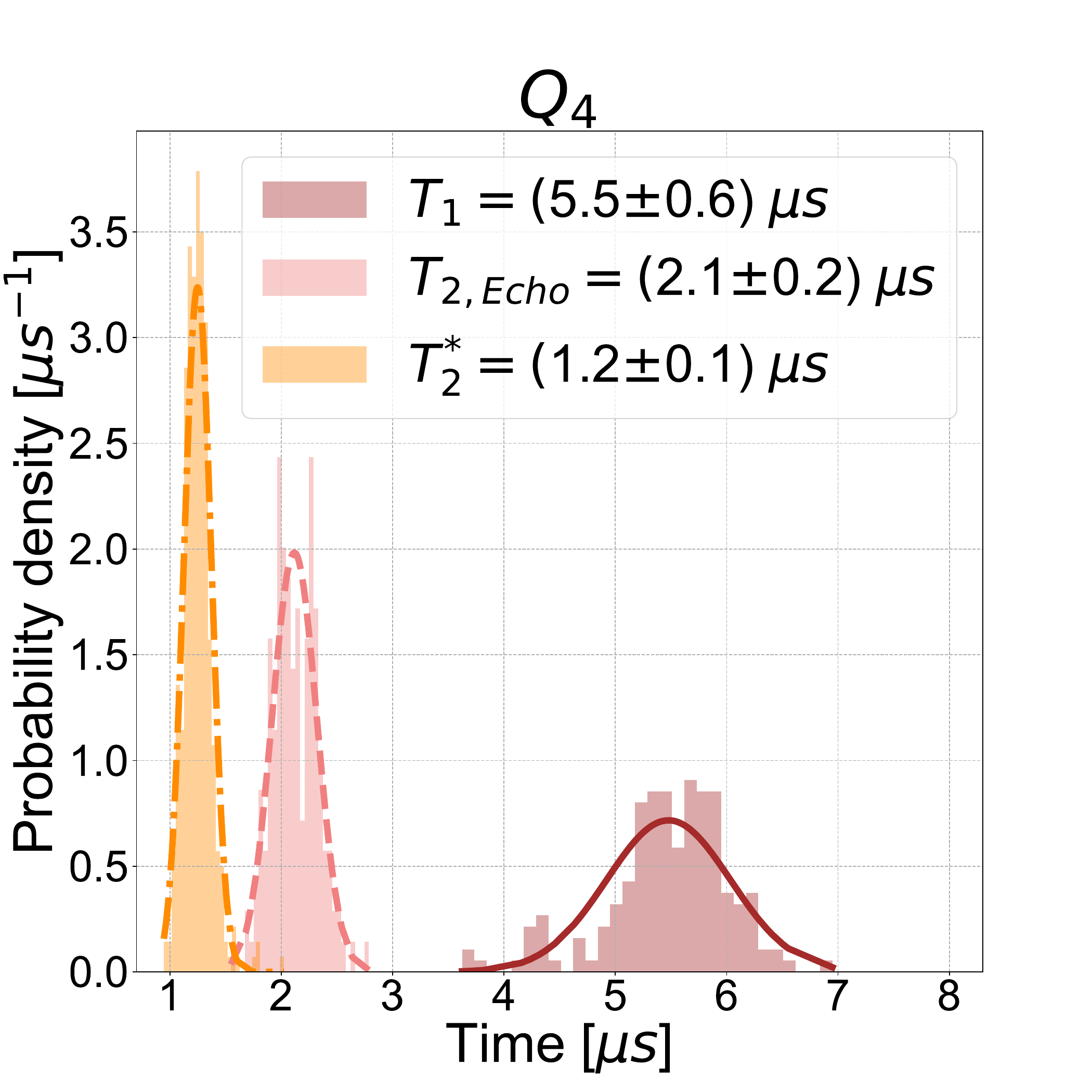}}
	\caption{Relaxation time $T_1$ and coherence times measured through Ramsey ($T_2^*$) and Hahn-Echo protocols ($T_{2,Echo}$) probability densities on: one low-frequency qubit ($Q_0$) in (a), the medium-frequency qubit ($Q_2$) in (b) and one high-frequency qubit ($Q_4$) in (c) of the five-qubit matrix. Solid, dashed, and dashed-dotted lines represent the normal probability density functions that best-fit $T_1$, $T_2$, and $T_{2, Echo}$ probability densities, respectively.}
	\label{coherence}     
\end{figure}

In Table~\ref{electrodynamics}, we report the electrodynamics parameters of the circuit, including: the qubits first-order transition frequencies $\nu_{01}$, the readout resonators frequency in the low-photon regime $\nu\ped{RO}$~\cite{Naghiloo2019}, the qubits charging energy $E\ped c$, which is proportional to the transmon anharmonicity \cite{koch2007}, the qubits relaxation times $T_1$ and coherence times $T_2^*$ and $T_2\api{Echo}$, measured through Ramsey interferometry and Hahn-Echo experiments~\cite{Naghiloo2019,krantz2019}, respectively. Relaxation, Ramsey and Hahn-Echo decay times have been repeatedly acquired in a time span of $12$ hours, in order to test the stability of the device coherence. The probability densities statistics for one low-frequency qubit ($Q_0$), the central qubit ($Q_2$), and one high-frequency qubit ($Q_4$) are reported in Figure~\ref{coherence}. The data for each qubit have been acquired by setting it near its flux Sweet Spot (SS) $(\sim 0.85 \Phi\ped{SS})$ while keeping the other qubits far-detuned from the transition frequency of the qubit under test, to avoid residual coupling effects~\cite{Majer2007}. The flux working point used for the diagnostic of the full $5$-qubits matrix corresponds to the maximum current settable by the room-temperature electronics available at the time of the experiment, and used to generate the magnetic flux field threading the flux-tunable qubits. Such experimental limitation could explain the relatively low values of Ramsey $T^*_2$ coherence times reported in Table~\ref{electrodynamics}.
\begin{table}
\centering
 \caption{Summary of the five-qubit quantum processor parameters acquired at $\sim0.85\Phi\ped{SS}$, where SS stands for flux sweet spot: the qubit transition frequency $\nu_{01}$, the charging energy $E\ped c$, the readout resonator frequency in the low-photon regime $\nu\ped{RO}$, the average qubit relaxation time $T_1$, Hahn-echo time ${T_2}\api{Echo}$, and Ramsey time ${T_2}^*$. The values highlighted with a $**$ are single-shot parameters. The error on the qubit frequency, estimated as the center of a Ramsey fringes experiment, is a maximum error related to the acquisition step as a function of the qubit-drive frequency. The error on the resonator frequency is a maximum error related to the readout frequency acquisition step.\label{electrodynamics}}
\begin{tabular}{@{}lllllll@{}}
\hline
Qubit & $\nu_{01}$ [GHz] & $E\ped c$ [MHz] & $\nu\ped{RO}$ [GHz]	& $T_1$ [$\mu s$] & $T_2\api{Echo}$ [$\mu s$] &  $T_2^*$ [$\mu s$]\\
\hline
$Q_0$ & $4.5546\pm0.0003$  & $340\pm2$ & $7.2500\pm0.0001$ & $24 \pm 5$ & $10\pm3$ & $2\pm1$\\
$Q_1$ & $4.2070\pm0.0005$ & $306\pm2$ & $7.5800\pm0.0001$  & $25\pm5$& $12\pm5$ & $0.76$ ($**$)  \\
$Q_2$ & $5.6503\pm0.0003$  & $274\pm2$ & $7.6585\pm0.0001$  & $8\pm1$ & $6\pm1$ & $1.1\pm0.1$\\
$Q_3$ & $6.2683\pm0.0005$ & $-$ & $7.84431\pm0.0001$  & $8\pm2$ & $6\pm1$& $1.25$ ($**$)  \\
$Q_4$ & $6.3892\pm0.0005$  & $359\pm2$ & $7.9957\pm0.0001$  & $5.5\pm0.6$ & $2.1\pm0.2$ & $1.2\pm0.1$\\
\hline
\end{tabular}
\end{table}
\begin{figure}[t]
\centering
	\subfloat[][\centering]{\includegraphics[width=0.4\columnwidth]{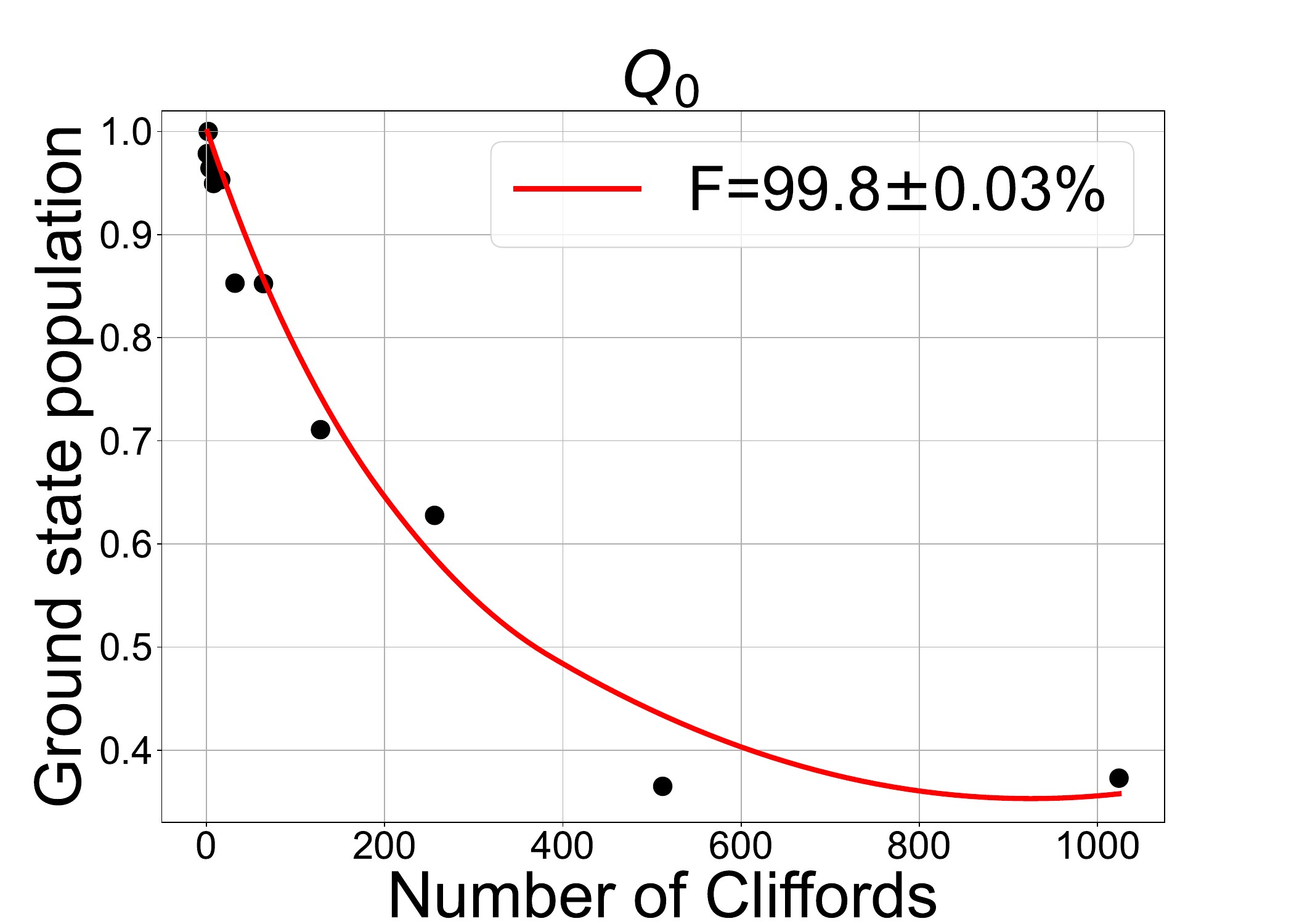}}
	\subfloat[][\centering]{\includegraphics[width=0.4\columnwidth]{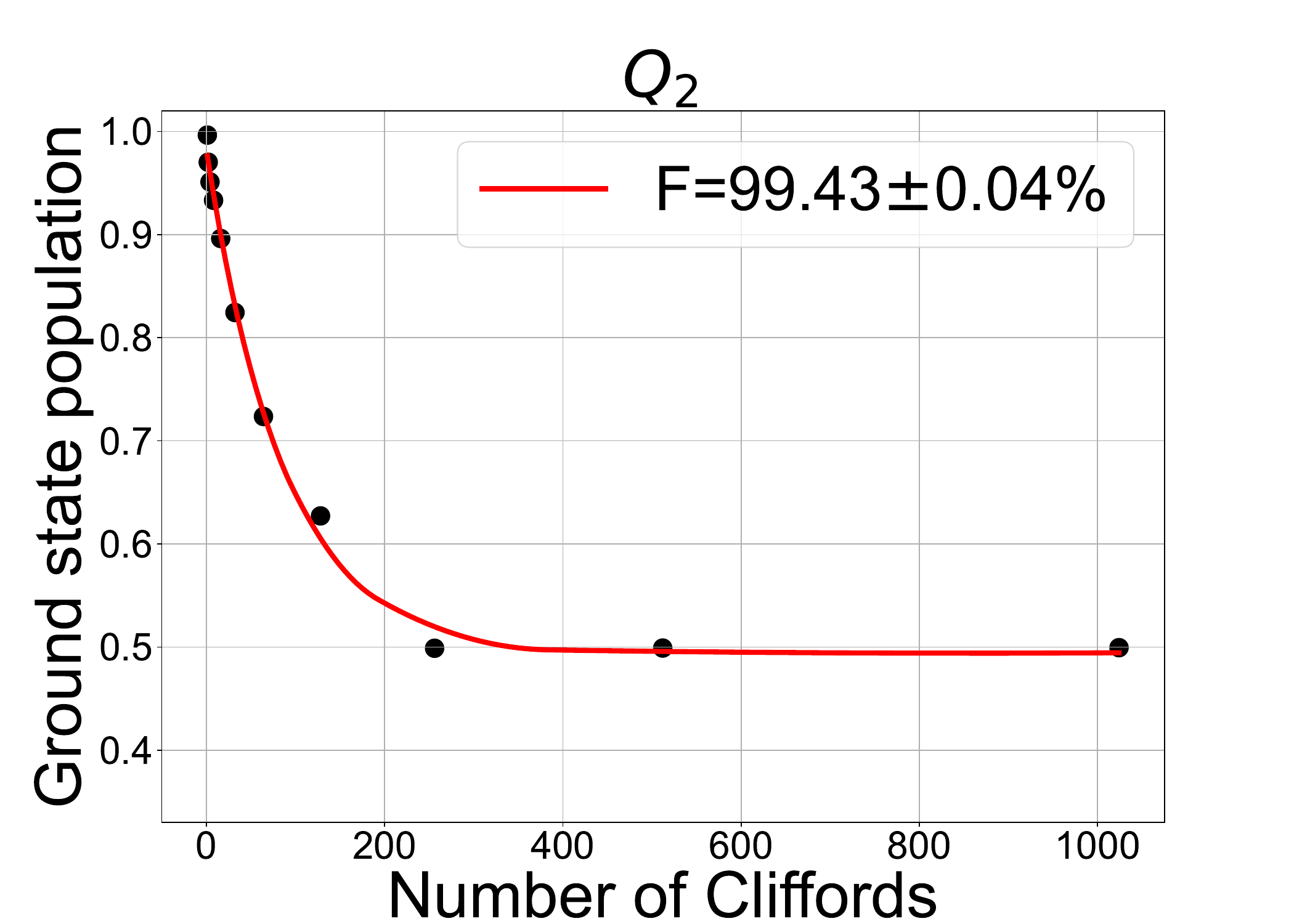}}\\
	\caption{Randomized benchmarking (RB) experiments for single-qubit gate fidelity estimation. In a) and b), black data points represent the ground state population as a function of the number of Clifford gates in the RB sequence for $Q_0$ and $Q_2$, respectively. The red lines are the power-law fit of the ground state population, which provides the average gate fidelity in the legends. For more information, please refer to the Supplementary Material.}
	\label{RB}     
\end{figure}    
\begin{figure}[t]	\includegraphics[width=0.5\columnwidth]
{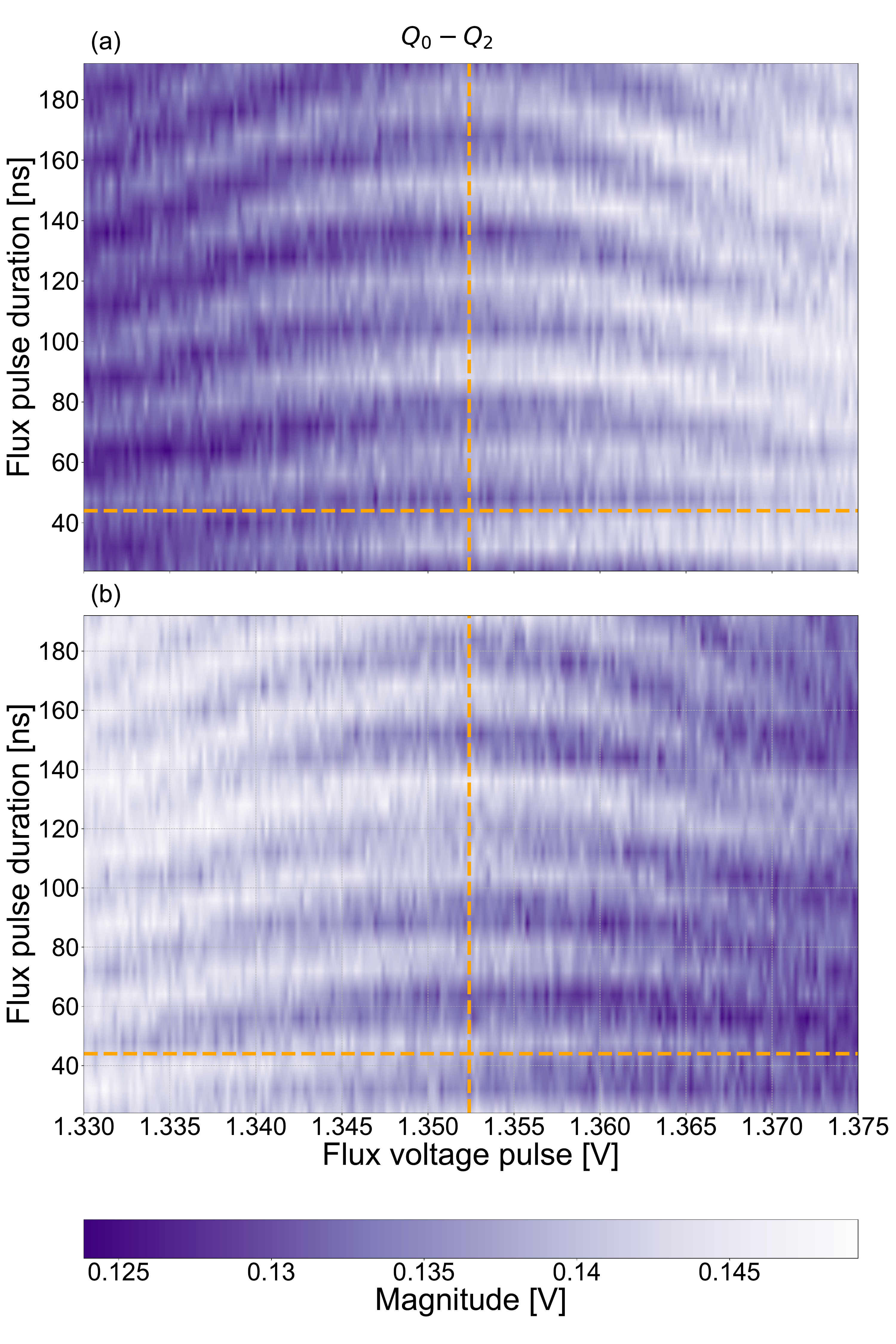}
\centering
	\caption{\emph{Chevron} plot for the pair $Q_0$-$Q_2$: on the y-axis, the CZ flux pulse duration; on the x-axis, the CZ flux pulse amplitude in voltage. The color scale represents the magnitude of the multiplexed readout signal: in (a), on the readout resonator of $Q_0$; in (b), on the readout resonator of $Q_2$. }
	\label{chevron}     
\end{figure}

Following the discussion from the previous section, to experimentally validate the FCM-QEM technique we must be able to initialize a quantum register and perform a set of quantum circuits made of single- and two-qubit gates. 
As a proof of principle, in this work, we validate the FCM-QEM performances focusing just on the pair $\{Q_0,Q_2\}$ of the $5$-qubit s-QPU, forming a $2$-qubit register. The choice to include $Q_2$ in the register comes from its central position in s-QPU, while the inclusion of a low-frequency qubit, rather than a high-frequency one, is related to the overall better coherence times of qubits from the former pair. For this specific set of experiments, we used an upgraded room-temperature experimental setup that allowed us to reach the flux SS of the qubits in the register. 

On this qubit subset, we perform all the operations required for the experimental characterization of the FCM-QEM technique. The quantum circuits employed for this purpose are based on specific combinations of single- and two-qubit gates, which we appropriately calibrated.
Single-qubit gates are performed through $20$ ns drive pulses applied on the qubits via dedicated drive lines. To reduce leakage and unwanted phase rotations, the drive pulses are DRAG (Derivative Reduction Adiabatic Gate) shaped ~\cite{Lucero2008,Motzoi2009,reed2013,Werninghaus2021,Babu2021}. To optimize the fidelity of the gate with respect to power, DRAG shape and frequency, we employ a combination of Motzoi calibration and All-XY technique~\cite{Motzoi2009,reed2013}. For more information on the calibration procedure, we refer to the Supplementary Material.

Within this single-qubit drive pulses optimization, we use Randomized Benchmarking (RB) experiments to measure the average single-qubit gate fidelity~\cite{Chow2010,sheldon2016,chao2018}. Results of the RB measurements are reported in Figure~\ref{RB}. Black data points represent the average ground state population in a RB experiment for $20$ randomization seeds as a function of the number of Clifford gates, while the red lines represent the fit of the ground state population, which provides a measure of the average gate fidelity $F$~\cite{Chow2010,sheldon2016,chao2018}. For more information on the fitting procedure, we refer to the Supplementary Material. Although manual iterative calibration of drive pulses allows to measure average gate fidelities $F=99.8\%$ for $Q_0$, fairly close to the state-of-the-art of $99.9\%$~\cite{sheldon2016, chao2018}, for $Q_2$ the best average gate fidelity is $F=99.43\%$.  

Two-qubit gates, such as the Conditional-Z (CZ) gate, are here built upon unipolar flux pulses~\cite{Strauch2003, DiCarlo2009}. Details on the coupling between each pair of the $5$-qubit s-QPU, i.\,e. on the avoided level crossing spectroscopy measurements, are reported in Supplementary Material. As for the optimization of CZ flux pulses on the $2$-qubit register $\ket{Q_0, Q_2}$, we report the so-known \emph{Chevron} plot~\cite{Negirneac2021} for $Q_0-Q_2$ pair in Figure ~\ref{chevron}. After setting the two qubits at their SS in the state $\ket{11}$, a unipolar flux pulse with variable amplitude and duration is applied on $Q_2$, via a dedicated flux line, in order to bring the states $\ket{11}$ and $\ket{20}$ of the 2-qubit register on resonance. The intersection between the dotted orange lines highlights the optimal values of the flux pulse duration and amplitude to implement a CZ gate, i.\,e. where the coherent exchange of energy between the states $\ket{11}$ and $\ket{20}$ is maximized~\cite{Negirneac2021}. We further calibrate the gate by performing conditional oscillation experiments around this optimal point~\cite{Rol2019}. To validate the fidelity of the two-qubit gate we perform two-qubit RB measurements. The results provide an average gate fidelity of $95\%$, while interleaving unipolar CZ gates in the RB sequence provides a CZ gate fidelity of the order of $\sim89\%$ (see Supplementary Material for additional information), well below typical average gate fidelities reported in the literature~\cite{Chen2014,Rol2019,Yuan2020,Sung2021,Negirneac2021,Marxer2023}. One reason for this is related to the fact that unipolar flux pulses have been demonstrated to be less efficient than other variants of the same gate that use Net-Zero (NZ) or Sudden-Net-Zero (SNZ) pulses, in terms of leakage~\cite{Negirneac2021}. 
\begin{figure}[t]	\includegraphics[width=0.5\columnwidth]
{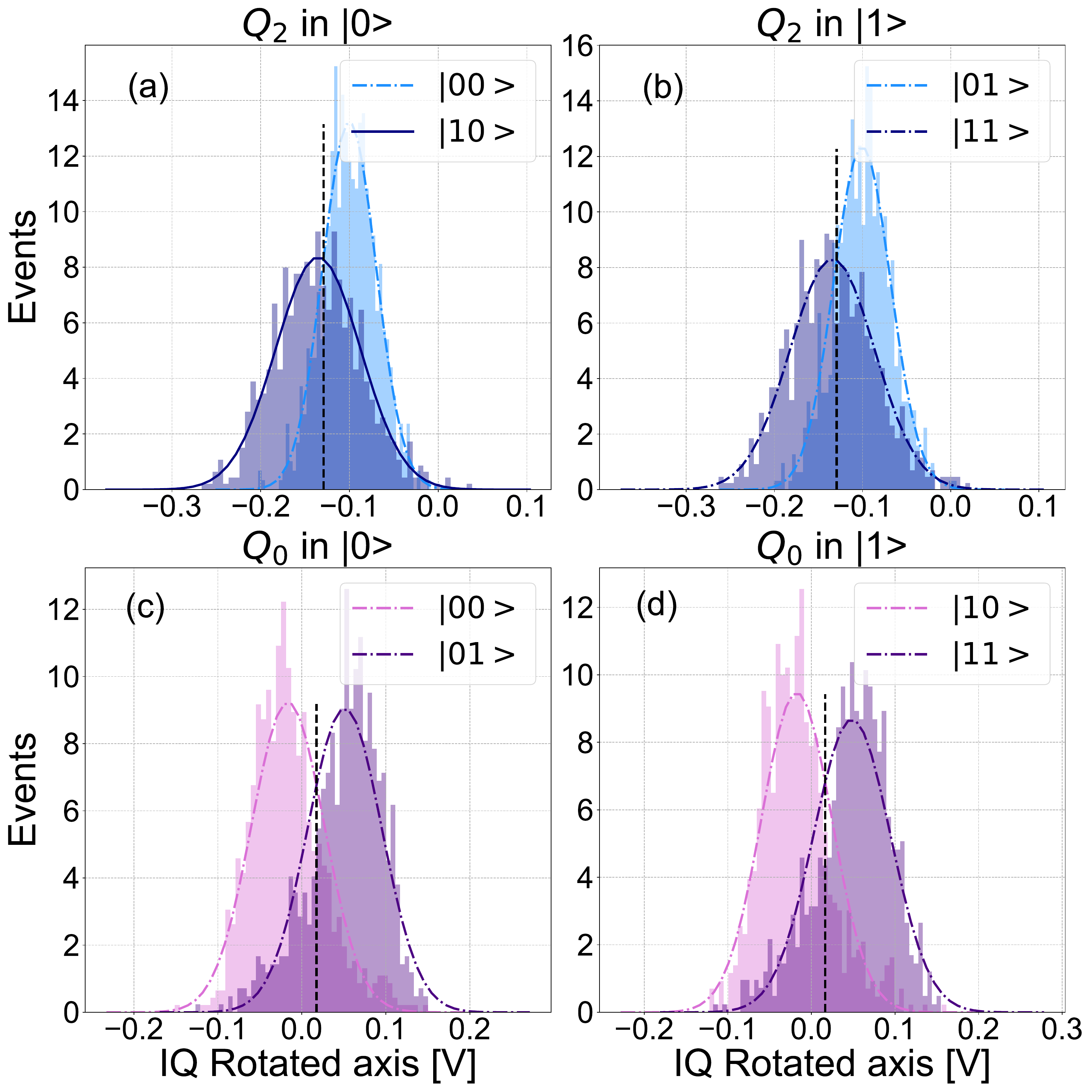}
\centering
	\caption{Two-qubits multiplexed I-Q gaussian distributions and readout threshold definition for state discrimination: in (a), for $Q_0$, when $Q_2$ is in $\ket{0}$; in (b), for $Q_0$, when $Q_2$ is in $\ket{1}$; in (c), for $Q_2$, when $Q_0$ is in $\ket{0}$ and in (d), for $Q_2$, when $Q_0$ is in $\ket{1}$.}
	\label{threshold}     
\end{figure}

Finally, since QEM techniques are meant to reduce the error of the expectation values of an observable, or more practically to mitigate readout errors on the output of quantum algorithms, it is important to report on the readout experimental procedure. The readout of the $2$-qubit register in the computational basis $\{\ket{00}, \ket{01},\ket{10},\ket{11}\}$ is performed by measuring a single-shot readout voltage signal $s=760$ times simultaneously on the readout resonators coupled to the two qubits~\cite{Mallet2009,Heinsoo2018,krantz2019,Chen2023}. The readout duration pulse is $200$ ns, and we here work in the dispersive and low-photon regime~\cite{krantz2019}, in order to provide Quantum Non-Demolition (QND) readout of the qubit state~\cite{krantz2019}. Specifically, the readout input signal power at the device is of the order of $-140$ dB. 

In the Supplementary Material, we show an example of experimental I-Q voltage distributions measured when preparing the two qubits in $\ket{0}$ and $\ket{1}$, where I is the in-phase (real) component of the readout voltage, and Q is the quadrature-phase (imaginary) component~\cite{krantz2019}. Due to the stochastic nature of the readout in NISQ quantum devices, the I-Q distributions follow a Gaussian behavior, where the mean value corresponds to the centroid of the \emph{I-Q blob}, and the standard deviation weights the readout spread and error~\cite{krantz2019}. An example of such Gaussian distributions is reported in Figure~\ref{threshold}. In order to determine the qubit state from the readout signal on the I-Q plane, we first rotate the complex values such that the entire readout signal is in only one quadrature (e.g., real), and then project each data point on the axis connecting the centroids of the ground and excited states Gaussian count distributions. If the projection falls above or below a specific threshold point (dashed lines in Figure~\ref{threshold}) on the connection axis, the qubit is in the ground or the excited state. Therefore, we quantify the readout probability $p_{ij}$ (where $i,j\in\{0,1\}$) to measure a specific state in the computational basis, after proper initialization in the same basis, by counting the number of events $n_{ij}$ for which the measured voltage falls above or below the threshold, as $p_{ij}=n_{ij}/s$. Specifically, the position of the threshold corresponds to the intersection between the two Gaussian distributions. If there is no intersection, we choose as readout threshold the midpoint between the two distributions. 

In Figure~\ref{threshold}, we compare the readout count distributions obtained through multiplexed simultaneous readout on $Q_0$ and $Q_2$ in their ground and excited states, while preparing the coupled qubit in $\ket{0}$ or $\ket{1}$. This is needed as a first experimental check on the quality of the final data. Indeed, the readout distributions measured on one qubit as a function of the state preparation should not depend on the state of the coupled qubit for non-entangled states, i.\,e. when applying only single-qubit gates in the circuit preparation. This is safely recovered in our experiments as well. 
\begin{figure}[t]
\centering
	\subfloat[][\centering]{\includegraphics[width=0.3\columnwidth]{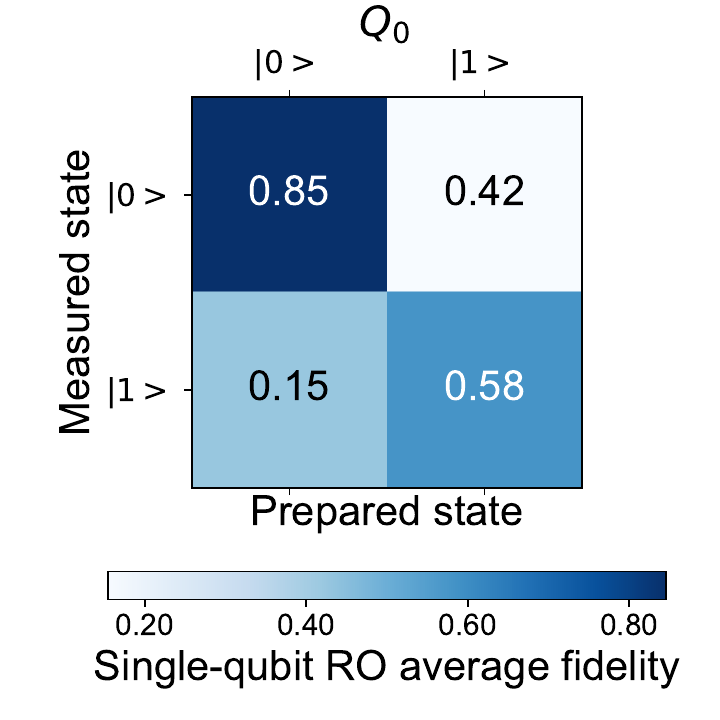}}
	\subfloat[][\centering]{\includegraphics[width=0.3\columnwidth]{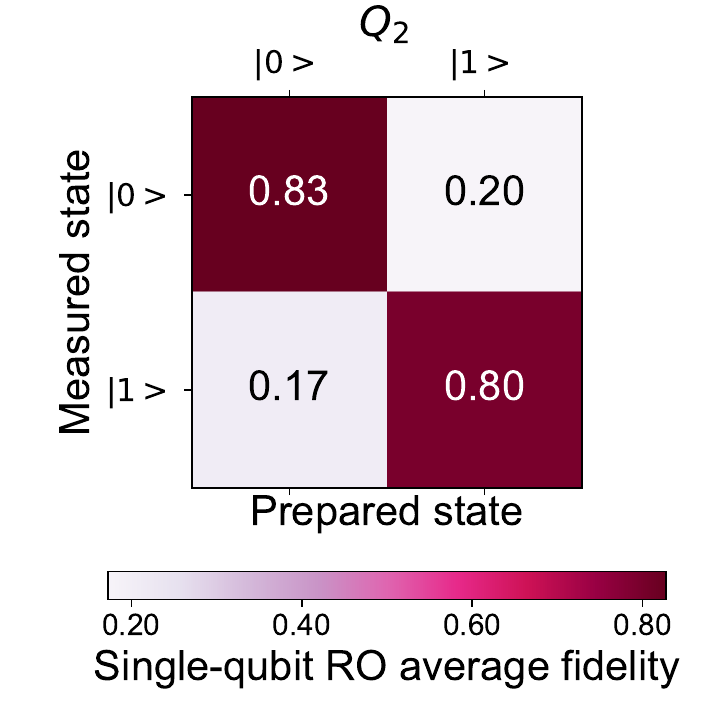}}\\
	\caption{Single-qubit readout fidelity (probability) matrix estimated with the threshold technique for $Q_0$ in (a) and for $Q_2$ in (b), on the basis $\{\ket{0},\ket{1}\}$.}
	\label{readoutfidelity}     
\end{figure}
In Figure~\ref{readoutfidelity}, we report the single-qubit readout probability matrix for $Q_0$ and $Q_2$ in terms of the prepared and measured states in the single-qubit basis $\{\ket{0}, \ket{1}\}$, when the other qubit is in the ground state. The readout fidelity for $Q_2$ is of the order of $80\%$, while for $Q_0$ this depends on the state preparation. It can range from $80\%$ when the qubit is prepared in $\ket{0}$ to $\sim60\%$ when it is excited. One of the possible reasons of the low readout fidelity achieved for $Q_0$ when prepared in the excited state, may be related to the presence of a spurious two-level system resonant with $Q_0$ frequency, which induces relaxation and dephasing~\cite{Degraaf2020}.

State-of-the-art superconducting quantum hardware experience readout errors from $20\%$ to a few percent, corresponding to readout fidelities of the order of $80\%$ up to $>90\%$. This is typically achieved by using quantum-limited amplification stages~\cite{Mallet2009,Reed2010,Riste2012,Schmitt2014,Walter2017}, or Purcell filtering~\cite{Sunada2022}, or through complex readout microwave techniques~\cite{Chen2023,DiPalma2023}. In this work, we stress that no quantum-limited amplification, nor sophisticated readout techniques have been adopted. 

Readout and gate fidelities, as well as the coherence performances of the device hereby presented, provide an ideal platform to test the FCM-QEM technique. To our knowledge, only a few works have proposed a study of different QEM techniques on superconducting quantum hardware~\cite{temme2017,kandala2019,Schulze2021,Cirstoiu2023}. Some of them took advantage of the several cloud platforms available online~\cite{Schulze2021,Cirstoiu2023}, which are characterized by readout fidelities above $94\%$, single-qubit gate fidelities reaching the golden standard of $99.9\%$, as well as two-qubit gate fidelities above $99.2\%$, and coherence times reaching hundreds of microseconds~\cite{temme2017,kandala2019,Schulze2021,Cirstoiu2023}. Remarkably, the system analyzed in this work allows us to identify a working regime where different quantum error patterns may play a significant role, thus fully exploiting the potential of the FCM-QEM technique.
\begin{figure}	
\subfloat[][\centering]{\includegraphics[width=0.65\columnwidth]{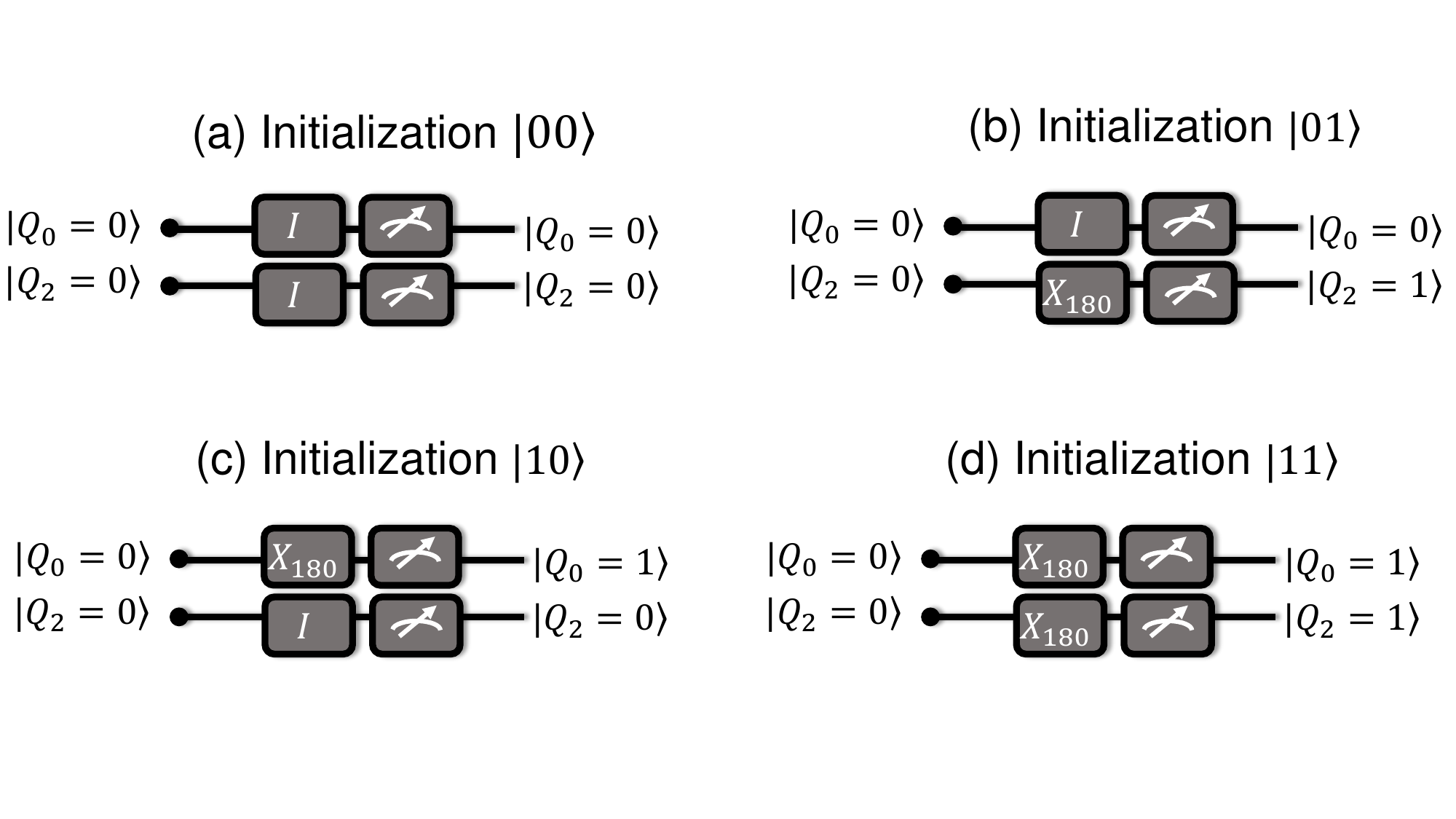}}
\subfloat[][\centering]{\includegraphics[width=0.35\columnwidth]
{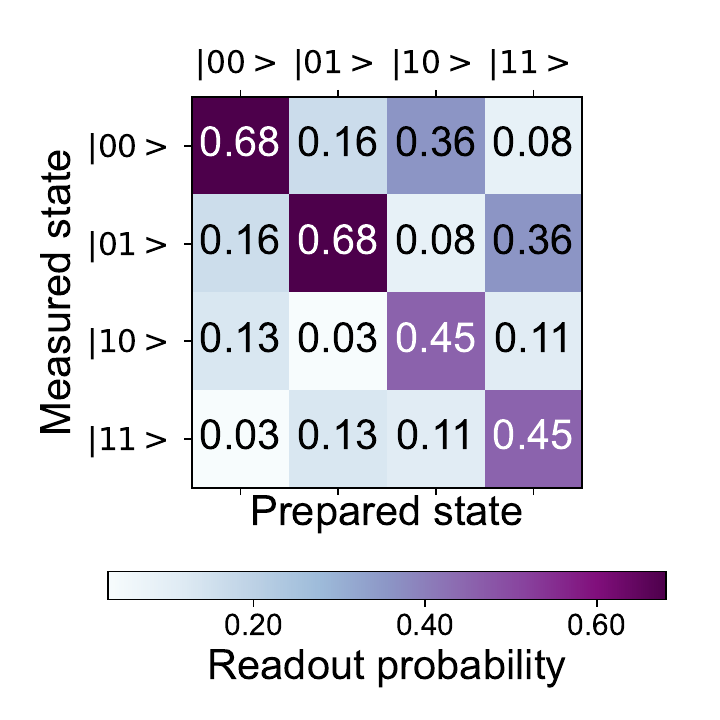}}
\centering
	\caption{In (a), initialization circuits; in (b), example of one initialization two-qubit readout fidelity (probability) matrix on the basis $\{\ket{00},\ket{01},\ket{10},\ket{11}\}$, where the first index state corresponds to $Q_0$ state, and the second to $Q_2$ state.}
	\label{initialization}     
\end{figure}

\section{Experiments and Results}
\label{sec:exp_and_res}

\subsection{Creation of the Datasets: Initialization Phase}
\begin{figure}	\includegraphics[width=1\columnwidth]
{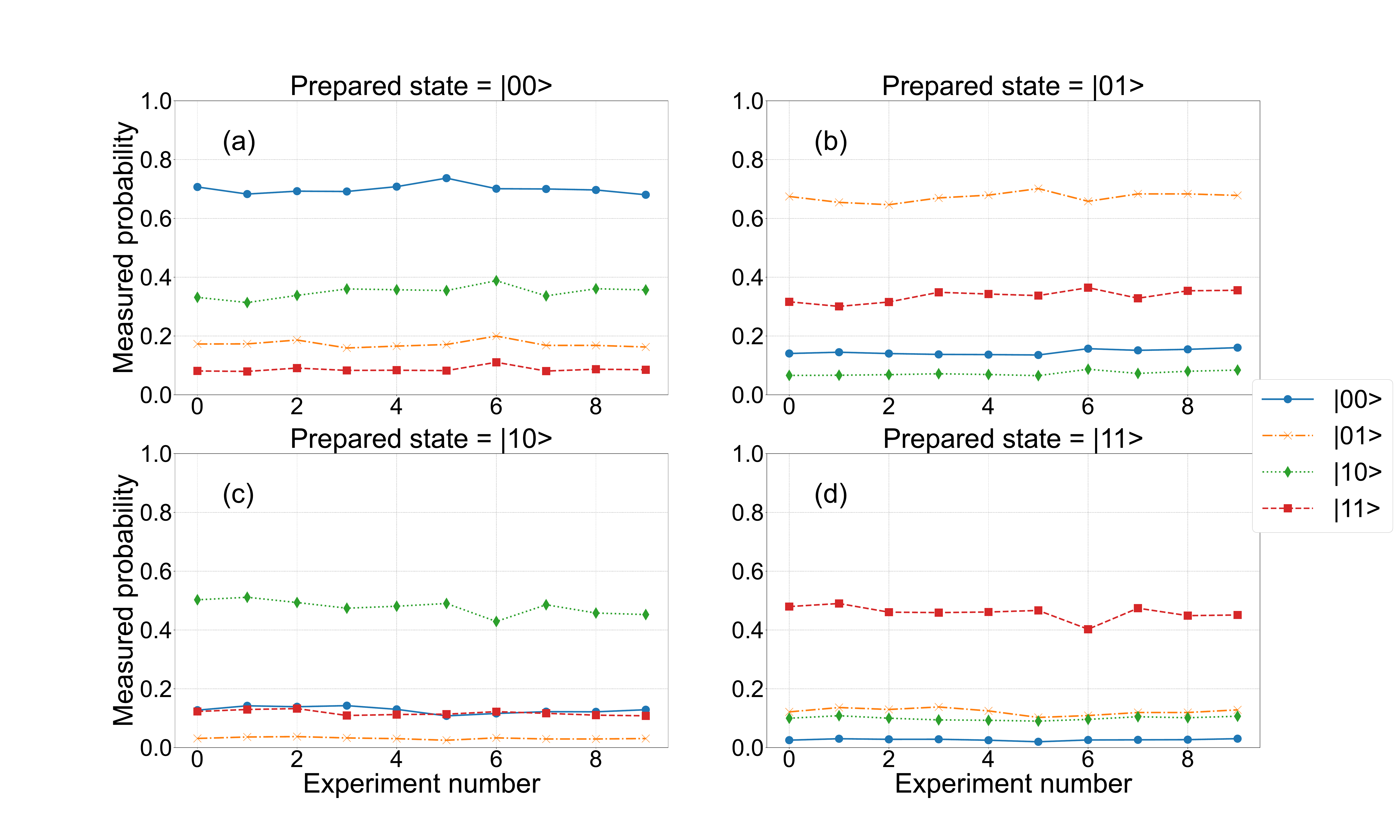}
\centering
	\caption{Readout probability over the $10$ initialization experiments for different measured states: in (a), for prepared state in $\ket{00}$, in (b), for prepared state in $\ket{01}$, in (c), for prepared state in $\ket{10}$, in (d), for prepared state in $\ket{11}$.}
	\label{stability}     
\end{figure}
The first step of the FCM-QEM algorithm (Section~\ref{sec:error_mitigation}) requires gathering information about the behavior of the s-QPU presented in Section~\ref{sec:hardware}. 
The $2$-qubit register $\ket{Q_0,Q_2}$ is initialized in each of the basis states $\{\ket{00}$, $\ket{01}$,$\ket{10}$,$\ket{11}\}$ multiple times ($t=10$) in order to collect the State Preparation And Measurement (SPAM) error information contained inside the initialization process. This means that four datasets, each made up of $t=10$ instances, are generated. In Figure~\ref{initialization} (a), we report the quantum circuits necessary to initialize the 2-qubit register in the basis states, where the gray boxes labeled with I represent the Identity gate, while the others represent qubit rotations around a certain axis. All single-qubit rotations on the Bloch sphere in the computational basis can be defined by the following unitary matrix:
\begin{equation}
\label{rotation}
R_{xy}(\theta,\phi)=\left(
\begin{matrix}
\cos\theta/2 & -i\sin\theta/2e^{-i\phi}\\
-i\sin\theta/2e^{i\phi} & \cos\theta/2\\
\end{matrix}
\right),
\end{equation}
where $\theta$ defines the rotation angle, while $\phi$ identify the phase of the rotation axis. For example, a rotation of $180 \degree$ around the x-axis is defined by the matrix $R_{xy}(\pi,0)$. In this work we just consider rotations around the x- and y-axis, thus we denote all single-qubit rotations with the reference axis of rotation $(X/Y)$ and the rotation angle. For each state preparation in Figure~\ref{initialization} (a), we measure the projection on the $2$-qubit register basis, and we calculate the two-qubit readout count probability as the tensor product between the readout probability matrix for $Q_0$, $p_{ij}^{{Q_0}}$, and $Q_2$, $p_{ij}^{{Q_2}}$, as $p_{ij}^{{Q_0}}\otimes{p_{ij}}^{{Q_2}}$. An example of one of the $10$ measured readout probability matrices is shown in Figure~\ref{initialization} (b).

Being the information on the readout error patterns in the $10$ datasets a footprint of the hardware noise-sensitivity, it is fundamental to avoid spurious inconsistencies in the output of the experiments that can alter the evaluation of the mitigation matrix. Unwanted variations in the hardware performances due to quantum effects and shot noise~\cite{Schulze2021} have been limited by running the $t=10$ iterations consecutively. The reliability of the datasets is shown in Figure~\ref{stability}, which represents the measured readout probability over the $t=10$ iterations, for each basis state preparation. The fluctuations in the readout probability are stable enough to guarantee that the different error patterns detected by the FCM-QEM algorithm are not related to different environmental conditions.

\subsection{Creation of the Mitigation Matrix}
The creation of the mitigation matrix $S$ for the $2$-qubit register involves the four datasets in Figure~\ref{stability}. According to~\cite{Acampora2021}, the hyper-parameters of the Fuzzy C-Means algorithm have been experimentally set as follows: $m = 2$, $maxiter = 10$, $\phi = 0.005$ and $C = 2,3,4$. 

The most appropriate probability vectors selected by FCM are used to compute the calibration matrix $M$ for the $2$-qubit register as follows:
\begin{equation}
    M = 
    \begin{pmatrix}
        0.74 & 0.16 & 0.36 & 0.08\\ 
        0.13 & 0.67 & 0.07 & 0.33\\
        0.11 & 0.03 & 0.48 & 0.12\\
        0.02 & 0.14 & 0.09 & 0.47
    \end{pmatrix}.
\end{equation}
Then, the mitigation matrix $S$ is given by the inverse of $M$, and can be implemented to correct the noisy outcomes of any quantum algorithm on the $2$-qubit register.

\subsection{Validation of the FCM-QEM Method}

\begin{figure}	\includegraphics[width=0.8\columnwidth]
{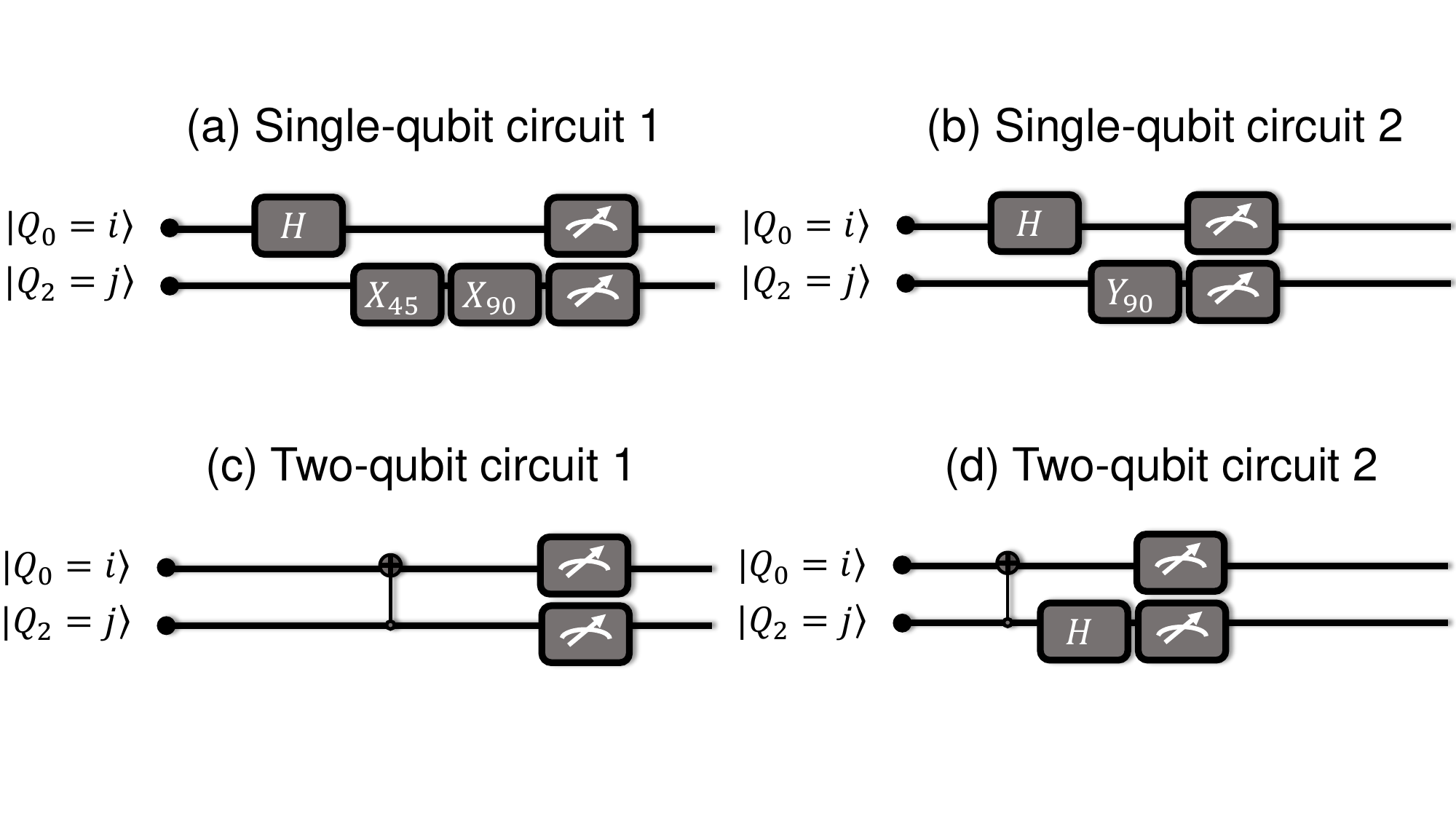}
\centering
	\caption{Single- and two-qubit gates quantum circuits used to validate the Fuzzy C-means quantum error mitigation technique, involving: in (a), one Hadamard gate and rotations around the X-axis of $45\degree$ and $90\degree$; in (b), one Hadamard gate and rotations around the Y-axis of $90\degree$; in (c), a CNOT-gate by using a CZ gate; in (d), one CNOT and a Hadamard gate. Each circuit has been implemented on different initial states $5$ times to guarantee a statistic on the error mitigation technique.}
	\label{circuits}     
\end{figure}

The experimental validation of the FCM-QEM method performed in this work consists of the implementation of four different test circuits. Each circuit is implemented after first initializing the register in all the possible basis states $\{\ket{00}, \ket{01},\ket{10},\ket{11}\}$, namely defining $16$ different test circuits. As shown in Figure~\ref{circuits}, the circuits in panels (a) and (b) consist only of single-qubit gates ($X_{45}$, $X_{90}$, $Y_{90}$), i.\,e. non-entangling gates, while the other two circuits in panels (c) and (d) include two-qubit gates, such as the CNOT. 

The $X_{45}$, $X_{90}$, and $Y_{90}$ gates follow the definition for single-qubit rotation in Eq.~\ref{rotation}. Specifically, a rotation of $90\degree$ around a specific axis is achieved by halving the amplitude of the $X_{180}$-pulse, while rotations of $45\degree$ by dividing the $X_{180}$-pulse by $4$~\cite{krantz2019}. The selection of the axis is provided by changing the phase $\phi$ of the control signals (Eq.~\ref{rotation})~\cite{krantz2019}. The Hadamard gate ($H$) is instead built upon single-qubit gate circuits combinations. Specifically, we disregard the role of the phase in the standard Hadamard matrix representation~\cite{krantz2019}, and we use a single-qubit $Y_{90}$ rotation followed by an $X_{180}$ gate~\cite{qinspire}. Finally, the CNOT gate in the two-qubit gate circuits in Figure~\ref{circuits} uses $Q_0$ as the target, since it shows a larger coherence time than $Q_2$, which here acts as the control qubit. Therefore, as usual, the CNOT is built applying first a Hadamard gate on the target followed by a CZ flux pulse, and another Hadamard on the target~\cite{krantz2019}.
\begin{figure}	\includegraphics[width=0.8\columnwidth]
{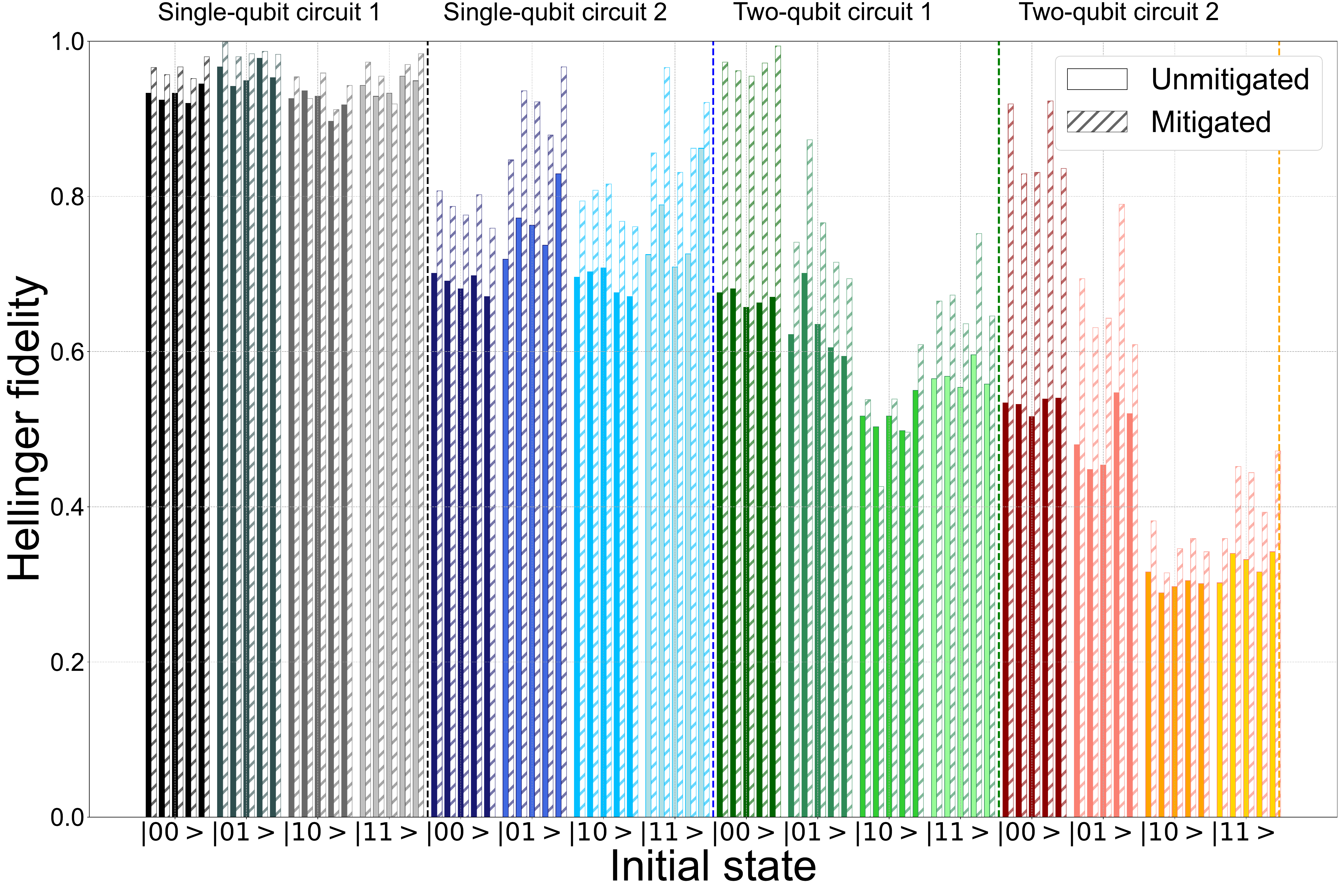}
\centering
	\caption{Comparison of the Hellinger fidelity as a function of the initial state preparation, for the unmitigated noisy circuits and the mitigated circuits for the $5$ experiments batches on the $4$ quantum circuits in Figure~\ref{circuits}, and as a function of the initial state preparation. }
	\label{mitigation}     
\end{figure}

In order to compare the unmitigated and mitigated outcomes against the noiseless (ideal) ones, we evaluate the Hellinger distance $H(P, Q)$~\cite{Hellinger1909}, a metric able to quantify the similarity between two discrete distributions $P=(p_1\dots p_k)$ and $Q=(q_1\dots q_k)$, given by
\begin{equation}
H(P,Q)=\frac{1}{2}\sqrt{\sum_{j=1}^k \left(\sqrt{p_j}-\sqrt{q_j}\right)^2}.
\end{equation}
Specifically, we compute the Hellinger fidelity, defined as $HF=(1-H^2(P, Q))^2$, between the ideal and the unmitigated probability distrubutions ($HF\ped{unmitigated}$), and between the ideal and the mitigated probability distrubutions ($HF\ped{mitigated}$). The higher the value of $HF$, the closer the mitigated/unmitigated probability distribution is to the ideal one. The experiments have been repeated $5$ times in order to collect statistical information on the performance of the mitigation matrix. 

In Figure~\ref{mitigation}, the obtained results are represented as blanck/dashed (unmitigated/mitigated) column bars for each quantum circuit and state preparation. For completeness, Table~\ref{hellinger} reports the average and the standard deviation of the Hellinger fidelity values over the $5$ repetitions, together with the corresponding improvement ($HF\ped{mitigated} - HF\ped{unmitigated}$) achieved through the FCM-QEM method.
\begin{table}[]
    \centering
    \begin{tabular}{c|c|c|c|}
         & $HF\ped{unmitigated}$ (\%) & $HF\ped{mitigated}$ (\%) & Improvement (\%) \\ \hline\hline
        \multicolumn{1}{c|}{Single-qubit circuit 1} &&& \\ 
        $\ket{00}$ & $93\pm1$& $96\pm1$& $+3\pm 1$\\
        $\ket{01}$ & $96\pm1$& $99\pm1$ & $+3\pm 1$ \\
       $\ket{10}$  &$92\pm2$& $94\pm2$ & $+2\pm 3$ \\
         $\ket{11}$& $94\pm1$ & $96\pm3$ & $+2\pm3$\\ \hline
        \multicolumn{1}{c|}{Single-qubit circuit 2} &&& \\ 
        $\ket{00}$&$69\pm1$& $79\pm2$& $+10\pm 2$ \\
        $\ket{01}$ & $76\pm4$& $91\pm 5$& $+15\pm 6$\\
        $\ket{10}$ &$69\pm 2$& $79\pm 2$& $+10\pm3$\\
        $\ket{11}$ & $76\pm6$& $89 \pm 5$& $+13\pm8$\\ \hline
        \multicolumn{1}{c|}{Two-qubit circuit 1} &&& \\ 
      $\ket{00}$  & $54\pm2$& $87\pm5$& $+33\pm5$\\
       $\ket{01}$ & $49\pm4$& $67\pm7$& $+18\pm8$\\
       $\ket{10}$  &$30\pm1$& $35\pm2$& $+5\pm2$\\
       $\ket{11}$ & $33\pm2$& $42\pm5$& $+9\pm5$\\ \hline
        \multicolumn{1}{c|}{Two-qubit circuit 2} &&& \\ 
     $\ket{00}$& $67\pm1$ & $97\pm1$& $+30\pm1$\\
    $\ket{01}$ & $63\pm4$& $76\pm7$& $+13\pm8$\\
        $\ket{10}$& $52\pm2$& $52\pm7$& $0\pm7$\\
        $\ket{11}$ & $57\pm2$& $67\pm5$& $+10\pm 5$\\ \hline
         \multicolumn{3}{r|}{\textit{Mean}}& $+11\pm 1$  \\
        \multicolumn{3}{r|}{\textit{Min}}& $+0$\\
        \multicolumn{3}{r|}{\textit{Max}}& $+33\pm 5$\\ \hline
    \end{tabular}
    \caption{Average Hellinger fidelity for the unmitigated and mitigated quantum circuits with the FCM-QEM technique in Figure~\ref{mitigation}, over the $5$ repetitions, improvement for each circuit and initial state, average, minimum, and maximum improvement. The errors for the Hellinger fidelities are given as the numerical standard deviation over the $5$ experiments, while the error on the improvement is obtained through standard error propagation.\label{hellinger}}
    \label{tab:my_label}
\end{table}

\section{Conclusion and Discussion}
\label{sec:conclusion}

The comparison of the Hellinger fidelity as a function of the preparation state, the implemented quantum circuits and the $5$ different runs between the unmitigated and mitigated output in Figure~\ref{mitigation} allows us to outline crucial conclusions about the performance of the FCM-QEM technique.

First, the Hellinger fidelity for the unmitigated outputs, i.\,e. on the NISQ device, depends on the quantum circuits themselves. The single-qubit circuit 1 has an average Hellinger fidelity ranging from $92$ to $96\%$. The single-qubit gate circuit 2 has a lower Hellinger fidelity, ranging  from $69$ to $76\%$. In both cases, there is no evident discrepancy among the different initial state preparations, and these values are in line with what is typically found on state-of-the-art devices~\cite{Schulze2021}. The inclusion of two-qubit gates lowers the Hellinger fidelity: for the two-qubit circuit 1, it is of the order of $63-67\%$, and sensitively decreases below $52-57\%$ when preparing $Q_0$, i.\,e. the target qubit, to $\ket{1}$. This is even more drastic when considering the CNOT alone, as in the two-qubit circuit 2 (see Table~\ref{hellinger}). As mentioned, this is intrinsically related to the quality of the CZ gate~\cite{Negirneac2021}. Improving the quality of the CZ gate may help in increasing the Hellinger fidelity on the NISQ device. 

Secondly, the CNOT gate alone in two-qubit circuit 2 seems to perform worse than the CNOT gate combined with a Hadamard gate. Although further investigation is required to quantitatively address this issue, one possible reason for it is related to the highest readout fidelity to measure the state of $Q_0$ in $\ket{0}$ when it is in $\ket{0}$ compared to the readout fidelity to measure the state $\ket{1}$ (Figure~\ref{readoutfidelity}). Therefore, this may suggest that a strong role is played by the quality of the readout in the device. For Hellinger fidelities in the unmitigated case lower than $\sim33\%$, the FCM-QEM technique improves the readout output of just $5-9\%$. Physically, this means that if a quantum circuit on a NISQ device completely fails, i.\,e. for Hellinger fidelities below $\sim50\%$, which corresponds to no similarity between the ideal and the unmitigated quantum circuits distributions~\cite{wadhia2023}, the error mitigation technique can not provide a sensitive correction to the output. This is, however, in line with other QEM techniques already reported. However, when the Hellinger fidelity is of the order of $60\%$, the FCM-QEM technique can provide an improvement on the readout up to about $30\%$. Even more noticeably, the FCM-QEM technique allows to increase the Hellinger fidelity to around $99-99.9\%$, even in best-case scenarios where the unmitigated circuit already shows Hellinger fidelity of the order of $~90\%$, as in the case of the single-qubit gate circuit 1. In ~\cite{Schulze2021}, the average improvement on the output of a three-qubit GHZ (Greenberger-Horne-Zeilinger) game~\cite{Vaidman1999}, provided by different QEM techniques on IBM-Q Athens and Yorktown, was also of the order of $\sim10-30\%$, demonstrating that the FCM-QEM a valid technique for QEM.

Finally, the repeatability of the FCM-QEM performance on several runs of the same quantum circuit, regardless of the preparation state, makes the technique sufficiently stable and allows to fully take into account the readout randomness and the shot noise of NISQ devices. 

In addition to the stability of the technique, the FCM-QEM has the advantage to be easily implementable, since it simply requires an initialization routine on the multi-qubit register prior to the algorithm meant to be run. This can be classically applied in post-processing, without necessarily overloading the already limited NISQ hardware capabilities during the algorithms, such as by correction gates in the algorithm, like in the ZNE~\cite{kandala2019,kim2023} or in the t-ReX QEM~\cite{vandenberg2022}.
On the other hand, as for the other QEM techniques that correct readout errors by applying a mitigation matrix \cite{vandenberg2022}, the FCM-based QEM is predicted to require exponential efforts to build and invert the aforementioned matrices when scaling up the number of qubits in the register. This problem will be addressed in future works. As for the time being, the quality of the readout improvement even in the experimental cases of not-fully optimized quantum circuits in this research, and the capability of the FCM-based QEM technique to be integrated on a software basis just simply adding to the interface architectures between the room-temperature hardware setup and the NISQ device, make this approach a suitable candidate to perform readout error mitigation on NISQ devices similar to the one used in this study. 

\section{Acknowledgements}
\label{sec:acknowledgements}

The work was supported by the project “SQUAD—On-chip control and advanced read-out for superconducting qubit arrays” Programma STAR PLUS 2020, Finanziamento della Ricerca di Ateneo, University of Napoli Federico II, the project SuperLink - Superconducting quantum-classical linked computing systems, call QuantERA2 ERANET COFUND, CUP B53C22003320005, the PNRR MUR project PE0000023-NQSTI and the PNRR MUR project CN\_00000013 -ICSC. G.A acknowledges financial support from the project PNRR MUR project PE0000013-FAIR.

H.G.A., D.M. (Davide Massarotti) and F.T. thank SUPERQUMAP project (COST Action CA21144), and the invaluable technical support provided by the electronics and software engineers at Qublox (Delft, Netherlands), and Orange Quantum Systems (Delft, Netherlands). 

\section{Conflict of Interest}

The authors declare no conflict of interest.

\section{Author Contributions}

\textbf{Halima G. Ahmad}, \textbf{Davide Massarotti} and \textbf{Francesco Tafuri}: Conceptualization (formulation of the approach for hardware and experimental setup); Supervision (physics activities); Writing-Original Draft Preparation.
\textbf{Alessandro Bruno}: Resources (quantum processor); Writing-Original Draft Preparation. 
\textbf{Giovanni Acampora} and \textbf{Autilia Vitiello}: Conceptualization (formulation of the approach for quantum mitigation error); Supervision (computer science activities); Methodology (design of the quantum mitigation error algorithm); Software (implementation of the quantum mitigation error algorithm); Writing–Original Draft Preparation. \textbf{Halima G. Ahmad}, \textbf{Pasquale Mastrovito} and \textbf{Anna
Levochkina}: Investigation (data collection), Methodology (implementation of the quantum circuits on the hardware and calibration of control, readout and two-qubit gates), Validation (analysis of mitigated results), Writing – Original Draft Preparation.
\textbf{Angela Chiatto} and \textbf{Roberto Schiattarella}: Investigation (testing activities with the quantum mitigation error algorithm), Validation (analysis of mitigated results); Writing–Original Draft Preparation. \textbf{Martina Esposito}, \textbf{Domenico Montemurro} and \textbf{Giovanni Piero Pepe}: Writing – Review \& Editing. \textbf{Francesco Tafuri}, \textbf{Giovanni Piero Pepe},\textbf{Giovanni Acampora}, and \textbf{Davide Massarotti}: Funding Acquisition. 
 
\section{Data availability statement}

Due to confidentiality agreements, the diagnostic data on the superconducting device supporting this study, so as the sample, can only be made available to personnel subject to a non-disclosure agreement. The code for the FCM-based QEM technique is available at:\\\href{https://github.com/Quasar-UniNA/FuzzyQMEM}{https://github.com/Quasar-UniNA/FuzzyQMEM}
\bibliographystyle{MSP}

\end{document}